\documentstyle[12pt,epsfig]{article}
\def\bm#1{\mbox{\boldmath $#1$}}

\def\ct#1{{\cal #1}}
\def\dfrac#1#2{{\displaystyle {#1 \over #2}}}
\newcommand{\ewxy}[2]{\setlength{\epsfxsize}{#2}\epsfbox[5 240 320 350]{#1}}
\newcommand{\nn}{\nonumber}

\newcommand{\Pj}{\mbox{I}\!\!\mbox{P}}
\newcommand{\Pjpv}{\mbox{1}\!\!\mbox{$^\wp$}}
\newcommand{\ctLambda}{\mbox{$\cal{J}$}\!\!\mbox{$\setminus$}}
\newcommand{\id}{\mbox{1$\!\!$I}}

\newcommand{\<}{\langle}
\renewcommand{\>}{\rangle}
\newcommand{\Tr}{\mbox{Tr}\;}

\newcommand{\QCD}{\mbox{\scriptsize QCD}}

\newcommand{\MSbar}{\overline{\mbox{MS}}}
\newcommand{\be}{\begin{equation}}
\newcommand{\ee}{\end{equation}}
\newcommand{\bea}{\begin{eqnarray}}
\newcommand{\eea}{\end{eqnarray}}

\begin{document}
\pagestyle{empty} 
%\begin{titlepage} \today \\
\begin{flushright}
EDINBURGH-97/11 \\
FTUV/99-4 \\
IFIC/99-4 \\
ROME1-1181/97 \\
ROM 2F/99/04 \\
\today
\end{flushright}
\centerline{\LARGE \bf Non-Perturbative Renormalization of}
\vskip 0.2cm
\centerline{\LARGE \bf Lattice Four-Fermion Operators without}
\vskip 0.2cm
\centerline{\LARGE \bf Power Subtractions}
\vskip 0.3cm
\centerline{\bf{A. Donini$^{a}$, V. Gim\'enez$^b$, G. Martinelli$^a$, 
M. Talevi$^c$, A. Vladikas$^{d}$}}
\vskip 0.3cm
\centerline{$^a$ Dip. di Fisica, Univ. di Roma ``La Sapienza'' and
INFN, Sezione di Roma,}
\centerline{P.le A. Moro 2, I-00185 Roma, Italy.}
\smallskip
\centerline{$^b$ Dep. de Fisica Teorica and IFIC, Univ. de Valencia,}
\centerline{Dr. Moliner 50, E-46100, Burjassot, Valencia, Spain}
\smallskip
\centerline{$^c$ Department of Physics \& Astronomy, University of Edinburgh}
\centerline{The King's Buildings, Edinburgh EH9 3JZ, UK}
\smallskip
\centerline{$^d$ INFN, Sezione di Roma II, and 
Dip. di Fisica, Univ. di Roma ``Tor Vergata'',}
\centerline{Via della Ricerca Scientifica 1, I-00133 Roma, Italy.}

\date{}
\abstract{
A general non-perturbative analysis of the renormalization properties of
$\Delta I=3/2$ four-fermion operators in the framework of lattice
regularization with Wilson fermions is presented. We discuss the
non-perturbative determination of the operator renormalization constants 
in the lattice Regularization Independent (RI or MOM) scheme. We also
discuss the determination of the finite lattice subtraction coefficients
from Ward Identities. We prove that, at large external virtualities,
the determination of the lattice mixing 
coefficients, obtained using the RI renormalization scheme, is equivalent to 
that based on Ward Identities, in the continuum and chiral limits.
As a feasibility study of our method, we compute the mixing 
matrix at several renormalization scales, for three values of the lattice
coupling $\beta$, using the Wilson and tree-level improved SW-Clover actions.
}
\vfill\eject
\pagestyle{empty}\clearpage
\setcounter{page}{1}
\pagestyle{plain}
%\date{}

%\end{titlepage}
\newpage
\pagestyle{plain} \setcounter{page}{1}

\section{Introduction}
\label{sec:introduction}

The renormalization constants of lattice
operators are necessary ingredients in the prediction of physical
amplitudes from lattice matrix elements. Schematically,
the physical amplitude $A_{\alpha \rightarrow \beta}$ associated to the 
transition $\alpha \rightarrow \beta$ induced by the composite operator $O$ 
is given, in the formalism of the Operator Product Expansion (OPE), by
\be
A_{\alpha \rightarrow \beta} = C_W(\mu) Z_O(a \mu)
\langle \alpha \vert O(a) \vert \beta \rangle 
\label{eq:ope}
\ee
where $C_W$ is the Wilson coefficient, $\mu$ is the renormalization scale, 
$a$ is the lattice spacing, $\langle \alpha \vert O (a)\vert \beta \rangle$
is the  matrix element of the bare  lattice operator relevant to the physical
process and $Z_O$ its  renormalization constant\footnote{
Here our notation is schematic: in general we have a set of operators and
$Z_O$ is a matrix which mixes those operators which form a complete basis
under renormalization.}.
The Wilson coefficient $C_W(\mu)$ is calculated in perturbation
theory at the renormalization scale $\mu$. It depends on the 
renormalization scheme (a mass-independent scheme is
implied throughout this paper). Wilson coefficients are known to NLO; see
refs.~\cite{altarelli}--\cite{ciucnlo}. The matrix element must
be calculated non-perturbatively; the only known method for computing it
from first principles (at a fixed cutoff $a^{-1}$)
is lattice QCD~\cite{CMPBGGM}.
The operator renormalization constant $Z_O$ is the link between
the matrix element, regularized on the lattice, and its
renormalized, continuum counterpart.
So far, three methods have been implemented to calculate it:
\begin{enumerate}
\item
Lattice Perturbation Theory (PT) \cite{msmith}- \cite{FREZZOTTIBORRELLI}.
\item
Lattice Ward Identities (WI), in the spirit of ref.~\cite{ks,BOCHICCHIO},
either with hadronic states  (see \cite{lucgui}-\cite{crluvl}), or with quark
states (see \cite{ZETA_A,Paciello} and, more recently \cite{jlqcd_lat96}),
or even with the Schr\"odinger functional (see \cite{lusch}). The WI method
is non-perturbative.
\item
The Non-Perturbative (NP) method of \cite{NP}, which consists in imposing
renormalization conditions at quark correlation functions with operator
insertions, at a given scale (see \cite{NP}-\cite{wustl}).
\end{enumerate}
Once the systematic
errors of these calculations are understood and kept under control, the matrix
elements, computed in simulations, can be properly renormalized and reliable
non-perturbative estimates of physical quantities can be obtained. 

The simplest case is that of the matrix elements of two-fermion
(dimension-three) operators, from which, for example, the decay constants of
light vector and pseudoscalar mesons can be extracted. Several complications
are avoided due to the fact that the operators renormalize multiplicatively.
The situation is less straightforward in the case of
four-fermion (dimension-six) operators. Their renormalization is also
additive; mixing occurs with other operators of the same dimension and, in some
cases, also with operators of lower dimensionality. These operators
must then be subtracted in order to make the original operator finite.
Mixing with lower-dimensional operators (e.g in the problem of understanding
the $\Delta I = 1/2$ rule) is characterized by several theoretical and
technical subtleties, which have been addressed in \cite{DI=1/2}.

This paper consists in a study of the renormalization properties of the 
complete basis of dimension-six four-fermion operators in the absence of
power subtractions (i.e. operators mixing only with others of equal dimension),
using non-perturbative methods. We will discuss below several
phenomenological applications of this problem, related to light (strange)
flavour phenomenology. We point out however, that since we are dealing with
a mass-independent renormalization scheme, our results can also be applied to
heavy-flavour phenomenology (e.g. the $\Delta B = 2$ process).
This calculation is also of relevance
to the evaluation of the $\Delta I = 1/2$ matrix elements, since the
anomalous dimension of the octet operators is unchanged by the mixing with
lower dimensional operators (and the related power subtractions)
\cite{sharperome}.
The most common examples, concerning $K$-meson decays, which 
involve operator mixing without power subtractions are the following
(note that SU(2)-isospin breaking effects are neglected):
\begin{itemize}
\item{
The study of $K^0$--$\bar K^0$ oscillations involves the computation
of $B_K$ (the B-parameter of the $K$-meson), which is obtained from
the $\Delta S = 2$ matrix element $\langle \bar K^0 \vert
O^{\Delta S = 2} \vert K^0 \rangle$ of the operator
\be
O^{\Delta S = 2}= ({\bar s} \gamma^L_{\mu} d)({\bar s} \gamma^L_{\mu}  d)
\label{ods2}
\ee
where $s$ and $d$ stand for strange and down quarks. Our conventions
for Dirac and colour matrices, indices etc. are given in
Appendix \ref{app:not}.
}
\item{
The study of the $\Delta I = 3/2$ sector of the decay $K \rightarrow \pi \pi$
involves the computation of the matrix elements $\langle \pi \pi \vert
O^{3/2}_{9,10} \vert K \rangle$ of the operators
\bea
O^{3/2}_{9} = 
({\bar s^A} \gamma^L_{\mu} d^A) ({\bar u^B} \gamma^L_{\mu}  u^B) +
({\bar s^A} \gamma^L_{\mu} u^A) ({\bar u^B} \gamma^L_{\mu}  d^B) -
({\bar s^A} \gamma^L_{\mu} d^A) ({\bar d^B} \gamma^L_{\mu}  d^B) \nn \\
O^{3/2}_{10} = 
({\bar s^A} \gamma^L_{\mu} d^B) ({\bar u^B} \gamma^L_{\mu}  u^A) +
({\bar s^A} \gamma^L_{\mu} u^B) ({\bar u^B} \gamma^L_{\mu}  d^A) -
({\bar s^A} \gamma^L_{\mu} d^B) ({\bar d^B} \gamma^L_{\mu}  d^A)
\label{eq:olli32}
\eea
with $u$ standing for the up quark and $A,B$ denoting colour indices. The 
above operators transform as members of the $(27,1)$ representation of
the $SU(3)_L \otimes SU(3)_R$ chiral group. Moreover,
the study of the electropenguin contribution to the decay
$K \rightarrow \pi \pi$ involves the computation of the matrix element
$\langle \pi \pi \vert O^{3/2}_{7,8} \vert K \rangle$ of the operators
\bea
O^{3/2}_{7} = 
({\bar s^A} \gamma^L_{\mu} d^A) ({\bar u^B} \gamma^R_{\mu}  u^B) +
({\bar s^A} \gamma^L_{\mu} u^A) ({\bar u^B} \gamma^R_{\mu}  d^B) -
({\bar s^A} \gamma^L_{\mu} d^A) ({\bar d^B} \gamma^R_{\mu}  d^B) \nn \\
O^{3/2}_{8} = 
({\bar s^A} \gamma^L_{\mu} d^B) ({\bar u^B} \gamma^R_{\mu}  u^A) +
({\bar s^A} \gamma^L_{\mu} u^B) ({\bar u^B} \gamma^R_{\mu}  d^A) -
({\bar s^A} \gamma^L_{\mu} d^B) ({\bar d^B} \gamma^R_{\mu}  d^A)
\label{eq:olri32}
\eea
which are also $\Delta I = 3/2$, but transform as an $(8,8)$ representation
of the chiral group. These
$K \rightarrow \pi\pi$ matrix elements are essential to the
calculation of $\epsilon^\prime/ \epsilon$. They can be obtained,
through soft pion theorems, from the matrix elements
$\langle \pi^+ \vert O^{3/2}_k \vert K^+ \rangle$ (for $k=7,8,9,10$). From
the single-state matrix elements,
the corresponding $B$-parameters, $B_{7,8}^{3/2}$, can be extracted.
Note that, in the limit of degenerate quark masses,
both operators $O^{3/2}_{9,10}$, having a ``left-left" Dirac
structure, renormalize in the same way as the $O^{\Delta S = 2}$ operator.
Moreover, the matrix elements
$\langle \pi^+ \vert O^{3/2}_{9,10} \vert K^+ \rangle$ have the same
$B$-parameter as $\langle \bar K^0 \vert O^{\Delta S = 2}
\vert K^0 \rangle$. These matrix elements should vanish in the chiral limit.
On the contrary, the operators $O^{3/2}_{7,8}$,
having a ``left -right" Dirac structure, obey different renormalization
properties. The corresponding matrix elements
$\langle \pi^+ \vert O^{3/2}_{7,8} \vert K^+ \rangle$ do not vanish
in the chiral limit
}
\item{
Important information on the physics beyond the Standard Model, such as SUSY,
can be obtained by studying FCNC processes and, in particular, $\Delta F = 2$
transitions (see \cite{Bsusy1} and references therein for a discussion).
Besides the $B$-parameters of the operators listed above, such
processes also require the knowledge of the $B$-parameters of the operators
\bea 
({\bar s^A} (1-\gamma_5) d^A) ({\bar s^B} (1-\gamma_5) d^B) \nn \\
({\bar s^A} (1-\gamma_5) d^B) ({\bar s^B} (1-\gamma_5) d^A) \nn \\
({\bar s^A} (1-\gamma_5) d^A) ({\bar s^B} (1+\gamma_5) d^B)
\label{eq:bsusy} \\
({\bar s^A} (1-\gamma_5) d^B) ({\bar s^B} (1+\gamma_5) d^A) \nn
\eea
}
\end{itemize}
Recent lattice results on all these $B$-parameters (with Wilson fermions)
can be found in refs.~\cite{jlqcd_lat96,gbs,b78910,Bsusy2}.

The matrix elements discussed above are extracted from the large time
asymptotic
behaviour of three-point correlation functions of the form $\langle P_K
O^{\Delta S = 2} P_K \rangle$ and $\langle P_K^\dagger O^{3/2}_k P_\pi
\rangle$ (for $k=7,8,9,10$) where $P_K$ and $P_\pi$ are pseudoscalar sources
of suitable quark flavour. Expressed in terms of traces of quark propagators, 
these correlation functions correspond to
the so-called ``eight"- shaped quark diagrams.
Any ``eye"-shaped contributions of these correlation functions cancel in the
limit of degenerate up and down quarks, which is the case under study.

%The present work consists in a study of the renormalization properties
%of the complete basis of dimension-six four-fermion operators, in the absence
%of power subtraction, using non-perturbative methods.
The paper is organized as follows: in sec.~\ref{sec:ds=2} we illustrate,
as a concrete example of the problems arising in lattice renormalization, the
mixing of the operator $O^{\Delta S = 2}$. Its matrix element
$\langle \bar K^0 \vert O^{\Delta S = 2} \vert K^0 \rangle$ is very sensitive
to the various systematic errors which affect its chiral behaviour and
therefore provides a good case-study. In sec.~\ref{sec:d=6} we discuss in full
detail the problem of mixing under renormalization of all four-fermion
dimension-six operators. The operators are classified according to their
discrete symmetries and an operator basis, convenient for our purposes,
is chosen. Subsequently, the operator mixing under renormalization is
divided into two parts: mixing that would occur even if all
the symmetries were respected by the regulator, and mixing which is
induced on the lattice by the chiral symmetry breaking of the Wilson term.
In sec.~\ref{sec:mixing} the NP renormalization is applied to the cases at
hand. The renormalization conditions of the so-called RI (or MOM) scheme are
worked out in terms of projected amputated Green functions, for the
complete operator basis\footnote{For continuum calculations in the RI scheme
and the anomalous dimensions of the complete basis of operators, calculated
in PT to NLO, see ref.~\cite{ciucnlo}.}. 
In sec.~\ref{sec:rcwi} we derive WI's suitable for
the determination of the lattice mixing coefficients in the chiral limit.
We show that, in this respect, the RI and the WI methods are equivalent in
the region of large external momenta. We also show the independence of the
lattice mixing coefficients from the renormalization scale.
Finally, in sec.~\ref{sec:res} we present the renormalization constants
and lattice mixing coefficients, obtained with the NP method,
with both the Wilson and the tree-level improved Clover action at several
values of the lattice bare coupling $\beta$. Some of these results have
been used in refs.~\cite{b78910,Bsusy2} in order to derive fully
non-perturbative estimates of the $B$-parameters.

\section{Renormalization of $O^{\Delta S = 2}$ and systematic errors}
\label{sec:ds=2}

The testing ground for the restoration of chiral symmetry in
the continuum limit of lattice computations has been the chiral behaviour of
the matrix element
$\langle \bar K^0 \vert O^{\Delta S = 2} \vert K^0 \rangle$, which,
if properly renormalized, vanishes when the $K$-meson becomes massless
\cite{cabigel}. Although early attempts with Wilson fermions
\cite{GAVELA}--\cite{GUPTAB} have given reasonable measurements of $B_K$,
they have been less successful in controlling this chiral
behaviour\footnote{$B_K$ has also been obtained with staggered
fermions \cite{SHARPE} mainly in the quenched approximation (see
ref.~\cite{Sharpe96} for a review). The (surviving) chiral symmetry
in the staggered fermion formalism ensures the vanishing
of the relevant matrix element in the chiral limit.}.
The root of the problem is the operator subtraction outlined above, which
we now discuss in some detail:
$O^{\Delta S =2}$ mixes with other operators $O_i$ of the same
dimension but with ``wrong naive chirality". Thus, the ($\mu$-dependent)
$K^0$--$\bar K^0$ matrix element of the
renormalized operator $\hat O^{\Delta S = 2}$ is given in
terms of the ($a$-dependent) bare matrix elements by:
\be
\langle \bar K^0 \vert \hat O^{\Delta S =2} (\mu) \vert K^0 \rangle = 
\lim_{a \to 0} \langle \bar K^0 \vert Z^{\Delta S=2}_0 (a\mu,g_0^2) \left[ 
O^{\Delta S =2}(a) + \sum_i Z_i^{\Delta S=2} (g_0^2) O_i (a) \right]
\vert K^0 \rangle
\label{eq:mix}
\ee
The overall renormalization constant $Z^{\Delta S=2}_0 (a\mu,g^2_0)$ 
is logarithmically divergent whereas the $Z_i^{\Delta S=2}(g^2_0)$'s are 
finite mixing  coefficients; $g_0^2(a)$ is the lattice bare coupling
(also expressed as $\beta = 6/g_0^2$). All $Z$'s can be calculated, at
least in principle, in perturbation theory (PT), provided that $\mu,a^{-1}
\gg \Lambda_{\QCD}$. The one-loop perturbative calculation of
the $Z$'s, both for the Wilson and Clover lattice
actions, has been performed in \cite{MARTIWdraper}-
\cite{FREZZOTTIBORRELLI}. Although the renormalized operator
$\hat O^{\Delta S =2} (\mu)$ is constructed so as
to have the correct chiral properties in the continuum limit, 
the implementation of operator mixing in early works
\cite{GAVELA}--\cite{GUPTAB} was subject to two main systematic errors:
\begin{enumerate}
\item
The overall renormalization constant $Z_0^{\Delta S=2}$ and the mixing 
coefficients $Z_i^{\Delta S=2}$ were calculated in one-loop PT.
Thus, they suffered from $\ct{O}(g_0^4)$ systematic errors. 
\item
The bare matrix elements
$\langle \bar K^0 \vert {O}^{\Delta S = 2} (a) \vert K^0 \rangle$
and $\langle \bar K^0 \vert {O_i} (a) \vert K^0 \rangle$ were
calculated non-perturbatively (numerically) with the Wilson action
at fixed cutoff $a$ (i.e.\  fixed coupling $g_0^2$). They were therefore
subject to $\ct{O} (a)$ systematic errors. 
\end{enumerate}
Both sources of systematic error may be considered responsible for the 
non-vanishing of the matrix element 
$\langle \bar K^0 \vert {\hat O}^{\Delta S = 2} \vert K^0 \rangle$
in the chiral limit. In order to reduce these errors, the following
remedies have been proposed:
\begin{enumerate}
\item
Boosted Perturbation Theory (BPT) should, according to \cite{LEPAGE},
improve the behaviour of the perturbative series of the renormalization
constants.
The systematic error due to the 1-loop truncation of the
BPT estimates of the $Z$'s remains $\ct{O} (g_0^4)$, but it is hoped
that it is smaller than the $\ct{O} (g_0^4)$ error of the original
PT expansion.
\item 
Discretization errors can be reduced by using improved actions and operators.
To this purpose, besides the Wilson action, the Clover action
\cite{SWHEATLIE} has also been
implemented in the calculation of the weak matrix elements
$\langle \bar K^0 \vert {O}^{\Delta S = 2} (a) \vert K^0 \rangle$
and $\langle \bar K^0 \vert {O_i} (a) \vert K^0 \rangle$.
Matrix elements of tree-level improved operators, calculated with the
tree-level improved Clover action, have discretization errors of
$\ct{O} (g_0^2 a)$, instead of $\ct{O} (a)$ as in the Wilson case\footnote{
Recently, it was shown that the non-perturbative Clover improvement,
proposed in \cite{lusch} reduces the above error to
$\ct{O} (a^2)$. We have not implemented this version of the Clover
action in the present work, because it involves more complicated operator
mixing.}.
\item
Using the non-perturbative (NP) method of \cite{NP,DS=2}, a more accurate
evaluation of the renormalization constants can be achieved. The method
consists in imposing renormalization conditions directly on quark four-point
Green functions with operator insertions. Higher-order contributions, including
non-perturbative effects, are taken into account. With this
method, the $Z$'s have the same discretization errors (due to the finiteness
of the lattice spacing) as the operator matrix elements we are ultimately
interested in. 
Another non-perturbative method consists in the determination of the
mixing coefficients from WI's on quark states \cite{jlqcd_lat96}.
We stress that the WI method can only be applied to the lattice mixing
coefficients; the overall renormalization $Z_0^{\Delta S = 2}(a\mu,g_0^2)$
can only be determined non-perturbatively from NP methods; e.g. those of
refs.~\cite{lusch,NP,DS=2}. Both WI and NP results are expected to have similar
discretization errors.
\end{enumerate}
The systematic errors of the operator renormalization
described by eq.~(\ref{eq:mix}) depend upon the method of calculation
of its various terms: if PT (or BPT) is used for the $Z$'s and
the Wilson action for the matrix elements, the renormalization is
good to $\ct{O} (g_0^4)$ and $\ct{O} (a)$. If PT (or BPT) is used for the
$Z$'s and the tree-level Clover action is used for the matrix elements, the
renormalization is good to $\ct {O}(g_0^4)$ and $\ct {O}(g_0^2 a)$. 
If NP (or WI) renormalization is used for the $Z$'s then the error is
$\ct {O} (a)$ or $\ct {O} (g_0^2a)$, depending on the action used
(Wilson or tree-level Clover).
Of course, the same considerations are valid for the renormalization  of
the operators of eqs.~(\ref{eq:olli32})--(\ref{eq:bsusy}).
\par
A study of the effect of each of these three improvements
had first been made in ref.~\cite{B_K}, in which a direct comparison
of the various errors in the computation of the $\Delta S = 2$
matrix element was carried out.
The main result of \cite{B_K} is that the chiral behaviour of the renormalized
$\langle \bar K^0 \vert {\hat O}^{\Delta S = 2} \vert K^0 \rangle$
is not improved significantly if BPT is used  for the calculation of
the mixing constants $Z^{\Delta S=2}_i$ and the Clover action
is implemented in the computation of the matrix elements\footnote{
For a different conclusion see ref.~\cite{lel}}.
What really makes a difference is the NP computation of the $Z$'s,
which was shown to improve significantly the chiral behaviour of 
the matrix element. This conclusion has been confirmed in \cite{jlqcd_lat96} 
with Wilson fermions and for several $\beta$ values. We stress that
in \cite{jlqcd_lat96} the non-perturbative computation of the mixing
$Z$'s was performed using WI's on quark states, as opposed to the
NP method of \cite{B_K}.
\par
In all early works, ref.~\cite{B_K} included, the renormalization 
of $O^{\Delta S =2}$ consisted in the mixing with only three dimension-six 
operators with the ``wrong naive chirality". This mixing had been explicitly
derived in \cite{MARTIWdraper}, using lattice PT at one-loop. However, by
applying the general symmetry arguments of \cite{BERNARD2} to the case
in hand, one can show that a fourth operator should be added in the sum of
eq.~(\ref{eq:mix}), as pointed out in refs.~\cite{jlqcd_lat96,wustl,GUPTA}.
This operator, absent at the one-loop level of PT, has a mixing constant which
is at least $\ct{O} (g_0^4)$. It gives a small (but non-negligible)
contribution
to the restoration of the chiral behaviour of the matrix element \cite{wustl}.
In refs.~\cite{wustl,b78910} and the present work, this operator, omitted in
ref.~\cite{B_K}, has been properly taken into account.

\section{Mixing of dimension-six four-fermion operators on the lattice}
\label{sec:d=6}

In this section, we introduce the basic notation and study the mixing
of generic four-fermion operators in the presence of explicit
chiral symmetry breaking, induced by the Wilson term. This
implies that mixing with operators with the ``wrong naive chirality" 
is allowed, even in the chiral limit.
Following \cite{BERNARD2} (see also \cite{GUPTA}), we
obtain the complete basis of dimension-six, four-fermion 
operators which mix under renormalization,
relying on general symmetry arguments based on the
vector flavour symmetry, which survives on the lattice.
In order to solve this problem, it is convenient to work
with massless fermions with four distinct flavours $\psi_f,\ f=1,\dots,4$.
The correlation functions arising from these operators do not contain
``eye"-shaped contributions and therefore their renormalization properties
are identical, up to trivial factors, to those of the physical $\Delta S=2$
and $\Delta I = 3/2$ operators considered in sect.~\ref{sec:introduction}
(recall that we are assuming SU(2)-isospin symmetry).
The introduction of ``light" masses (i.e. masses which are much smaller
than the renormalization scale) will be
discussed in subsect.~\ref{subsec:scale}.
We will not address the problem of ``heavy" masses (of the order of, or
greater than the typical scales and the UV cutoff) in this work.

We define the generic four fermion operators
\begin{eqnarray}
O_{\Gamma^{(1)}\Gamma^{(2)}} &=& 
(\bar\psi_1\Gamma^{(1)}\psi_2)(\bar\psi_3\Gamma^{(2)}\psi_4)
\nonumber \\
O_{t^a \Gamma^{(1)} t^a \Gamma^{(2)}} &=& 
(\bar\psi_1 t^a \Gamma^{(1)} \psi_2)(\bar\psi_3 t^a \Gamma^{(2)} \psi_4)
\nonumber \\
O^F_{\Gamma^{(1)}\Gamma^{(2)}} &=& 
(\bar\psi_1\Gamma^{(1)}\psi_4)(\bar\psi_3\Gamma^{(2)}\psi_2)
\nonumber \\
O^F_{t^a \Gamma^{(1)} t^a \Gamma^{(2)}} &=& 
(\bar\psi_1 t^a \Gamma^{(1)} \psi_4)(\bar\psi_3 t^a \Gamma^{(2)} \psi_2)
\label{eq:qgam1gam1}
\end{eqnarray}
where $\Gamma^{(1)}$ and $\Gamma^{(2)}$ denote any Dirac matrices and $t^a$
are the colour-group generators (referred to as colour matrices in this work).
For notation and conventions, see Appendix \ref{app:not}.
\begin{table}
\centering
\begin{tabular}{|r|r|r|r|r|r|}
\hline
$O_{\Gamma^{(1)} \Gamma^{(2)}}$ & $\ct{P}$ & $\ct{CS}'$ & $\ct{CS}''$ &
$\ct{CPS}'$ & $\ct{CPS}''$ \\ \hline \hline
$O_{VV}$ & $+1$ & $+1$ & $+1$ & $+1$ & $+1$ \\
$O_{AA}$ & $+1$ & $+1$ & $+1$ & $+1$ & $+1$ \\
$O_{PP}$ & $+1$ & $+1$ & $+1$ & $+1$ & $+1$ \\
$O_{SS}$ & $+1$ & $+1$ & $+1$ & $+1$ & $+1$ \\
$O_{TT}$ & $+1$ & $+1$ & $+1$ & $+1$ & $+1$ \\
\hline
$O_{VA}$ & $-1$ & $-1$ & $-O_{AV}$  & $+1$ & $O_{AV}$  \\
$O_{AV}$ & $-1$ & $-1$ & $-O_{VA}$  & $+1$ & $O_{VA}$  \\
$O_{SP}$ & $-1$ & $+1$ & $O_{PS}$  & $-1$ & $-O_{PS}$  \\
$O_{PS}$ & $-1$ & $+1$ & $O_{SP}$  & $-1$ & $-O_{SP}$  \\
$O_{T \tilde T}$ & $-1$ & $+1$ & $+1$  & $-1$ & $-1$  \\
\hline
$O_{[VA+AV]}$ & $-1$ & $-1$ & $-1$ & $+1$ & $+1$ \\
$O_{[VA-AV]}$ & $-1$ & $-1$ & $+1$ & $+1$ & $-1$ \\
$O_{[SP-PS]}$ & $-1$ & $+1$ & $-1$ & $-1$ & $+1$ \\
$O_{[SP+PS]}$ & $-1$ & $+1$ & $+1$ & $-1$ & $-1$ \\
$O_{T \tilde T}$ & $-1$ & $+1$ & $+1$  & $-1$ & $-1$  \\
\hline
\end{tabular}
\caption{Classification of four- fermion operators
$O_{\Gamma^{(1)} \Gamma^{(2)}}$ according to $\ct{P}$ and useful
products of their discrete symmetries $\ct{C}$, $\ct{S'}$ and $\ct{S''}$. 
These properties are
also valid for the operators $O_{t^a \Gamma^{(1)} t^a \Gamma^{(2)}}$.
For the operators $O^F_{\Gamma^{(1)} \Gamma^{(2)}}$
and $O^F_{t^a \Gamma^{(1)} t^a \Gamma^{(2)}}$, we must
exchange the entries of the columns $\ct{CS}' \leftrightarrow \ct{CS}''$
and $\ct{CPS}' \leftrightarrow \ct{CPS}''$. Note that $O_{\tilde T \tilde T}
= O_{TT}$ and $O_{T \tilde T} = O_{\tilde T T}$.}
\label{tab:g12}
\end{table}

\subsection{Operator classification according to discrete symmetries}
\label{subsec:cpsbases}

Let us start by addressing the problem of mixing of the generic
operators of eq.~(\ref{eq:qgam1gam1}) under renormalization.
First of all we stress that they cannot mix with operators of lower
dimensionality, because such operators do not
have the four-flavour content of the original ones
\footnote{This statement remains true for the operators of
eqs.(\ref{ods2})--(\ref{eq:bsusy}).
The $\Delta I = 1/2$ operator however, mixes in general
with operators of dimension-five and -three.}.
Thus, $O_{\Gamma^{(1)}\Gamma^{(2)}}$
can mix with any other dimension-six operator, provided it has
the same quantum numbers
(i.e.\   with any operator which has the symmetries of 
$O_{\Gamma^{(1)} \Gamma^{(2)}}$ and of the action).
The generic QCD Wilson lattice action with 4 degenerate quarks
is symmetric under parity $\ct{P}$, and charge conjugation $\ct{C}$.
Moreover, there are three other useful (flavour) symmetries of the action,
namely the flavour exchange symmetry $\ct{S} \equiv (\psi_2
\leftrightarrow \psi_4)$ and the switching symmetries $\ct{S}' \equiv
(\psi_1 \leftrightarrow \psi_2 , \psi_3 \leftrightarrow \psi_4)$ and
$\ct{S}'' \equiv
(\psi_1 \leftrightarrow \psi_4 , \psi_2 \leftrightarrow \psi_3)$
\cite{BERNARD2}. In Table \ref{tab:g12} we classify the operators
$O_{\Gamma^{(1)} \Gamma^{(2)}}$, for all $\Gamma^{(1)}$ and $\Gamma^{(2)}$
combinations of interest, according to the discrete symmetries
$\ct{P}$, $\ct{C}$, $\ct{S}'$ and $\ct{S}''$.
Parity violating operators (middle section of the Table), for which
$\ct{CS}''$ is not a symmetry, have been (anti)symmetrized in order to
obtain eigenstates of $\ct{CS}''$ (bottom section of the Table).
We adopt the notation
\be
O_{[\Gamma^{(1)}\Gamma^{(2)} \pm \Gamma^{(2)}\Gamma^{(1)}]} = 
O_{\Gamma^{(1)}\Gamma^{(2)}}\pm O_{\Gamma^{(2)}\Gamma^{(1)}}
\ee
Note that the results of Table \ref{tab:g12} apply also to the
operators $O_{t^a \Gamma^{(1)} t^a \Gamma^{(2)}}$ since,
upon performing the symmetry transformations, sign differences,
resulting from the presence of the colour $t^a$ matrix, disappear because
the colour matrices appear quadratically.
On the other hand, $O^F_{\Gamma^{(1)} \Gamma^{(2)}}$ is
obtained by applying $\ct{S}$ on $O_{\Gamma^{(1)} \Gamma^{(2)}}$.
Since $\ct{S}$ transforms $\ct{S}'$ into $\ct{S}''$, the properties
of Table \ref{tab:g12} also apply to
$O^F_{\Gamma^{(1)} \Gamma^{(2)}}$, but with all $\ct{S}'$
and $\ct{S}''$ columns exchanged. Again, the operator
$O^F_{t^a \Gamma^{(1)} t^a \Gamma^{(2)}}$ has the same properties
as $O^F_{\Gamma^{(1)} \Gamma^{(2)}}$, since the colour matrix
$t^a$ appears quadratically.

Our aim is to find complete bases of operators which mix under renormalization.
Thus, the first task is the elimination of the operators which are not
independent. In particular, given the mixing
of $O_{\Gamma^{(1)} \Gamma^{(2)}}$ with all other allowed
$O_{\Gamma^{(i)} \Gamma^{(j)}}$'s and
$O^F_{\Gamma^{(i)} \Gamma^{(j)}}$'s,
we do not need to consider also the mixing with the
$O_{t^a \Gamma^{(i)} t^a \Gamma^{(j)}}$'s and
$O^F_{t^a \Gamma^{(i)} t^a \Gamma^{(j)}}$'s, since they can be expressed
in terms of the $O_{\Gamma^{(i)} \Gamma^{(j)}}$'s and the
$O^F_{\Gamma^{(i)} \Gamma^{(j)}}$'s. This can be seen by
applying the standard identity of colour matrices (Fierz transformation of
colour indices)
\be
t^a_{AB} t^a_{CD} = -\frac{1}{2N_c} \delta_{AB} \delta_{CD}
+\frac{1}{2} \delta_{AD} \delta_{CB}
\label{eq:tt}
\ee
on the $t^a$'s of a given operator. 
For the operator $O_{t^a \Gamma^{(i)} t^a \Gamma^{(j)}}$
the result has the general form
\be
O_{t^a \Gamma^{(i)} t^a \Gamma^{(j)}} = - \frac{1}{2N_c}
O_{\Gamma^{(i)} \Gamma^{(j)}} + \frac{1}{2} \sum_{n,m}
C_{nm} O^F_{\Gamma^{(n)} \Gamma^{(m)}}
\ee
where the sum runs over all the Dirac matrices obtained by the
Fierz transformation of $\Gamma^{(i)} \Gamma^{(j)}$. The factors
$C_{nm}$ are the appropriate constants of the Fierz transformation of the
Dirac matrices (see Appendix \ref{app:Fierz} for details on Fierz
transformations in Dirac space).
Analogously we can express $O^F_{t^a \Gamma^{(i)} t^a \Gamma^{(j)}}$
in terms of $O_{\Gamma^{(i)} \Gamma^{(j)}}$ and
$O^F_{\Gamma^{(i)} \Gamma^{(j)}}$.
Therefore, in the following,  it is adequate to limit ourselves to the
mixing of $O_{\Gamma^{(i)} \Gamma^{(j)}}$'s and
$O^F_{\Gamma^{(i)} \Gamma^{(j)}}$'s, 
which form a complete basis of 20 independent operators (these are the
5 operators of the top section of Table~ \ref{tab:g12}, the 5 operators
of the bottom section and their 10 Fierz counterparts).

Furthermore, by classifying the operators according to the discrete
symmetries listed in Table \ref{tab:g12}, the original basis of 20
operators can be further decomposed into smaller independent bases.
An immediate decomposition is that into two bases, of 10 operators each,
with definite parity ($\ct{P} = \pm 1$). Further decompositions occur
upon using the remaining $\ct{CPS'}$, $\ct{CPS''}$ and $\ct{S}$ symmetries.
These we now perform case by case.

\subsection{Parity conserving operators}
\label{subsec:pcbases}

All of the parity conserving operators $O_{\Gamma \Gamma}$ 
are eigenstates of all the discrete symmetries listed above, with
eigenvalue $+1$. Therefore, each of them can mix with the other four,
and also with the five $O^F_{\Gamma \Gamma}$'s; the complete
basis consists of 10 operators.
We now rotate our basis into a new one,
consisting of 10 operators which are also eigenstates of $\ct{S}$
with eigenvalues $\pm 1$:
\begin{equation}
O^\pm_{\Gamma \Gamma} = \frac{1}{2} \left [
O_{\Gamma \Gamma} \pm O^F_{\Gamma \Gamma} \right ] = \frac{1}{2}
\left [ (\bar\psi_1\Gamma\psi_2)(\bar\psi_3\Gamma\psi_4) \pm  
(\bar\psi_1\Gamma\psi_4)(\bar\psi_3\Gamma\psi_2) \right ]
\label{eq:qgamgam}
\end{equation}
Clearly, the 5 $O^+_{\Gamma \Gamma}$'s, corresponding to $\ct{S}=+1$, mix
only among themselves; the same is true for the 
$O^-_{\Gamma\Gamma}$'s which have $\ct{S}=-1$. Thus the original 
basis of 10 operators has been decomposed into two independent
bases of 5 operators each.

In this work, since we are eventually interested in the renormalization
of the operators of eqs.~(\ref{ods2})--(\ref{eq:bsusy}),
we opt for the basis
\bea
Q^\pm_1 & \equiv & O^\pm_{[VV+AA]} \nonumber \\
Q^\pm_2 & \equiv & O^\pm_{[VV-AA]} \nonumber \\
\label{eq:qq}
Q^\pm_3 & \equiv & O^\pm_{[SS-PP]} \\
Q^\pm_4 & \equiv & O^\pm_{[SS+PP]} \nonumber \\
Q^\pm_5 & \equiv & O^\pm_{TT} \nonumber
\eea
Other choices of basis can be found in \cite{MARTIWdraper,jlqcd_lat96}.
All such bases are simply linear combinations of each other.

\subsection{Parity violating operators}
\label{subsec:pvbases}

The parity violating four- fermion operators listed at the bottom section of
Table \ref{tab:g12} do not all have identical $\ct{CPS}'$ and $\ct{CPS}''$
values. We will now establish their mixing pattern, with the aid of $\ct{S}$
symmetry.

The operator $O_{[VA + AV]}$ mixes with $O^F_{[VA + AV]}$ only, forming a 
basis of two operators with $\ct{CPS}'= \ct{CPS}'' = +1$. In analogy to the
parity conserving case, we rotate this basis into
\be
O^\pm_{[VA+AV]} \equiv \frac{1}{2} \left [  O_{[VA + AV]}  \pm
O^F_{[VA + AV]} \right ]
\label{vaav24}
\ee
Since $O^+_{[VA+AV]}$ has $\ct{S} = +1$ and $O^-_{[VA+AV]}$ has $\ct{S} = -1$,
they do not mix with each other; the two operators renormalize multiplicatively.
In other words, the original basis of two operators has been decomposed into
two bases of one operator each.

The operators $O_{[VA - AV]}$ and $O^F_{[SP - PS]}$ mix, since they both have
$\ct{CPS}' = +1$ and $\ct{CPS}'' = -1$. Similarly, $O^F_{[VA - AV]}$ and
$O_{[SP - PS]}$ form another basis, with $\ct{CPS}' = -1$, $\ct{CPS}'' = +1$.
It is convenient to combine these two bases into 4 operators:
\bea
O^\pm_{[VA-AV]} & \equiv & \frac{1}{2} \left [(O_{[VA - AV]}
 \pm  O^F_{[VA - AV]} \right ]
\nonumber \\
O^\pm_{[SP-PS]} & \equiv & \frac{1}{2} \left [ O_{[SP - PS]} \pm
O^F_{[SP - PS]} \right ]
\label{sppsm24}
\eea
None of these operators have definite $\ct{CPS}'$ or $\ct{CPS}''$.
However, they have definite $\ct{S} = \pm 1$. 
There is mixing between the two $\ct{S} = +1$ operators ($O^+_{[VA-AV]}$,
$O^+_{[SP-PS]}$) and also between the two $\ct{S} = -1$ operators
($O^-_{[VA-AV]}$, $O^-_{[SP-PS]}$).

We follow a similar line of reasoning for the operators
$O_{[SP + PS]}$, $O^F_{[SP + PS]}$, $O_{T \tilde T}$ and $O^F_{T \tilde T}$
(all have $\ct{CPS}'=\ct{CPS}'' = -1$). We rotate the basis of these
four operators into
\bea
\nonumber
O^\pm_{[SP+PS]} & \equiv & \frac{1}{2} \left [  O_{[SP + PS]}  \pm
O^F_{[SP + PS]} \right ] \\
O^\pm_{T \tilde T} & \equiv & \frac{1}{2} \left [  O_{T \tilde T}  \pm
O^F_{T \tilde T} \right ]
\label{spps24}
\eea
Thus, once more,
the original basis of the four operators of eq.(\ref{spps24}) can be
decomposed into two bases, of two operators each, with
definite $\ct{S} = \pm 1$.

For reasons of notational compactness, we will use the following
redefinitions:
\bea
\ct{Q}^\pm_1 & \equiv & O^\pm_{[VA+AV]} \nonumber \\
\ct{Q}^\pm_2 & \equiv & O^\pm_{[VA-AV]} \nonumber \\
\label{eq:hh}
\ct{Q}^\pm_3 & \equiv & - O^\pm_{[SP-PS]} \\
\ct{Q}^\pm_4 & \equiv & O^\pm_{[SP+PS]} \nonumber \\
\ct{Q}^\pm_5 & \equiv & O^\pm_{T \tilde T} \nonumber
\eea

\subsection{Operator subtraction and chiral symmetry}
\label{subsec:chiralsub}

This completes the discussion of the mixing of dimension-six operators
on the lattice,
in the general case of 4 distinct flavours (degenerate or not).
We now summarize the result to keep in mind. For the
parity conserving operators, the renormalization is given by
\be
\label{eq:mrenpc}
\hat Q_i^\pm = Z^\pm_{ij} Q_{j}^\pm \qquad (i,j = 1,\dots ,5)
\ee
whereas for the parity violating ones we have
\be
\label{eq:mrenpv}
\hat {\ct{Q}} _i^\pm = \ct{Z}^\pm_{ij} \ct{Q}_j^\pm \qquad (i,j = 1,\dots ,5)
\ee
where $\hat Q _i^\pm$ and $\hat{\ct{Q}}_i^\pm$ are the bases of
renormalized operators and $Z_{ij}^\pm$ and $\ct{Z}_{ij}^\pm$
are the renormalization matrices (summation over repeated indices
is implied). 

On the basis of $\ct{CPS}$ symmetries,
it is important to notice that the matrix $\ct{Z}_{ij}^\pm$ is a
(relatively sparse) block diagonal matrix. We show this
explicitly by re-writing the last equation according to the allowed mixing
of the operators derived in the previous Subsection:
\begin{equation}
\left(\begin{array}{c} 
\hat{\ct{Q}}_1 \\
\hat{\ct{Q}}_2 \\
\hat{\ct{Q}}_3 \\
\hat{\ct{Q}}_4 \\
\hat{\ct{Q}}_5
\end{array}\right)^\pm
=
\left(\begin{array}{rrrrr}
\ct{Z}_{11} & 0 & 0 & 0 & 0 \\
0 & \ct{Z}_{22} & \ct{Z}_{23} & 0 & 0 \\
0 & \ct{Z}_{32} & \ct{Z}_{33} & 0 & 0 \\
0 & 0 & 0 & \ct{Z}_{44} & \ct{Z}_{45} \\
0 & 0 & 0 & \ct{Z}_{54} & \ct{Z}_{55}
\end{array}\right)^\pm
\left(\begin{array}{c} 
\ct{Q}_1 \\
\ct{Q}_2 \\
\ct{Q}_3 \\
\ct{Q}_4 \\
\ct{Q}_5
\end{array}\right)^\pm
\label{eq:renpv}
\end{equation}
The abbreviated notation with the $\pm$ superscript of the column vectors
and  matrix should be transparent to the reader.
In conclusion, the lattice does not induce extra subtractions for the parity
violating sector ($\ct{Q}_k^\pm$; $k=1,\dots,5$), since the mixing occurring
according to the pattern of eq.~(\ref{eq:renpv}), is also valid in the
$\chi RS$ scheme; see below.

In the case of parity conserving operators, we proceed in a different way.
We find it convenient to separate the above
operator mixing into two classes: 1) the first is the lattice subtraction,
which consists in correcting the operator mixing, induced by the breaking of
chiral symmetry due to the Wilson term of the action; 2) the second is the
subtraction which survives in the continuum limit.
In order to facilitate this separation,
let us suppose that there is a regularization scheme which, unlike
the lattice, respects chiral symmetry; i.e. the regularized action
has no chiral symmetry breaking term (recall we are working at zero quark
mass). We will call
this hypothetical scheme the Chirally-symmetric Regularization Scheme
($\chi RS$ for short). In this scheme, we can use chiral symmetry in
order to establish some extra selection rules for the renormalization of
the operators of interest. Although the resulting ``operator renormalization"
is incomplete on the lattice, it is the one which should be recovered
in the the continuum limit, since the renormalized theory must have the
desired chiral properties. The remaining lattice subtractions are
due to the presence of the Wilson term. 

In order to derive the selection rules in the $\chi RS$ scheme,
it is adequate to consider two discrete axial symmetries. 
The first (denoted by $\chi_{24}$) acts only on flavours 2 and 4:
\bea
&& \psi_2 \rightarrow i\gamma_5 \psi_2 \qquad ;
\qquad \bar\psi_2 \rightarrow i \bar\psi_2\gamma_5
\nonumber \\
&& \psi_4 \rightarrow i\gamma_5 \psi_4 \qquad ;
\qquad \bar\psi_4 \rightarrow i \bar\psi_4\gamma_5
\eea
The second symmetry (denoted as $\chi_{12}$) acts only on flavours 1 and 2:
\bea
&& \psi_1 \rightarrow  i\gamma_5 \psi_1 \qquad ;
\qquad \bar\psi_1 \rightarrow  i \bar\psi_1\gamma_5
\nonumber \\
&& \psi_2 \rightarrow  i\gamma_5 \psi_2 \qquad ;
\qquad \bar\psi_2 \rightarrow  i \bar\psi_2\gamma_5
\eea
Under these transformations, the four-fermion operators transform as shown
in Table~\ref{tab:chisym}.
\begin{table}
\centering
\begin{tabular}{|r|r|r|r|r|r|r|r|r|r|r|}
\hline
{\rm {Symmetry}} & $Q^+_1$ & $Q^+_2$ & $Q^+_3$ & $Q^+_4$ & $Q^+_5$ &
$Q^-_1$ & $Q^-_2$ & $Q^-_3$ & $Q^-_4$ & $Q^-_5$
\\ \hline \hline
$\chi_{24}$ & $-1$ & $+1$ & $+1$ & $-1$ & $-1$ &
$-1$ & $+1$ & $+1$ & $-1$ & $-1$ \\
\hline
$\chi_{12}$ & $+1$ & $Q^-_2$ & $-Q^-_3$ & $-1$ & $-1$ &
$+1$ & $Q^+_2$ & $-Q^+_3$ & $-1$ & $-1$ \\
\hline
\end{tabular}
\caption{Classification of four-fermion parity conserving operators
$Q^\pm_k$ ($k=1,\dots,5$), according to the discrete
symmetries $\chi_{24}$ and $\chi_{12}$. The parity violating operators
$\ct{Q}_k^\pm$ ($k=1,\dots,5$) transform in the same way.}
\label{tab:chisym}
\end{table}
The symmetry $\chi_{12}$ implies that $Q^\pm_1$ renormalize
multiplicatively, whereas $Q^+_4$ and $Q^+_5$ ($Q^-_4$  and $Q^-_5$) mix
 with each other.
From the $\chi_{24}$ symmetry we deduce that $Q^+_2$ and $Q^+_3$  
($Q^-_2$ and $Q^-_3$) also
mix with each other. In the absence of explicit chiral symmetry
breaking, we conclude that the  mixing structure is  the same
as that considered above for the parity-violating counterparts. 
The corresponding parity conserving operators ($Q_k^\pm$; $k=1,\dots,5$),
belonging to the same chiral representations, would have a $\chi RS$
mixing pattern similar to that of eq.~(\ref{eq:renpv}):
\begin{equation}
\left(\begin{array}{c} 
\hat Q_1 \\
\hat Q_2 \\
\hat Q_3 \\
\hat Q_4 \\
\hat Q_5
\end{array}\right)^\pm
=
\left(\begin{array}{rrrrr}
Z_{11} & 0 & 0 & 0 & 0 \\
0 & Z_{22} & Z_{23} & 0 & 0 \\
0 & Z_{32} & Z_{33} & 0 & 0 \\
0 & 0 & 0 & Z_{44} & Z_{45} \\
0 & 0 & 0 & Z_{54} & Z_{55}
\end{array}\right)^\pm
\left(\begin{array}{c} 
\tilde Q_1 \\
\tilde Q_2 \\
\tilde Q_3 \\
\tilde Q_4 \\
\tilde Q_5
\end{array}\right)^\pm
\label{eq:renor_subt}
\end{equation}
where the $\tilde Q_i$ represent the bare operators
in the $\chi RS$ scheme. 
In the presence of the Wilson term, the $\tilde Q_i$ are
the lattice subtracted operators defined as 
\begin{equation}
\left(\begin{array}{c} 
\tilde Q_1 \\
\tilde Q_2 \\
\tilde Q_3 \\
\tilde Q_4 \\
\tilde Q_5
\end{array}\right)^\pm
=
\left(\begin{array}{c} 
Q_1 \\
Q_2 \\
Q_3 \\
Q_4 \\
Q_5
\end{array}\right)^\pm
+
\left(\begin{array}{rrrrr}
0 & \Delta_{12} & \Delta_{13} & \Delta_{14} & \Delta_{15} \\
\Delta_{21} & 0 & 0 & \Delta_{24} & \Delta_{25} \\
\Delta_{31} & 0 & 0 & \Delta_{34} & \Delta_{35} \\
\Delta_{41} & \Delta_{42} & \Delta_{43} & 0 & 0 \\
\Delta_{51} & \Delta_{52} & \Delta_{53} & 0 & 0
\end{array}\right)^\pm
\left(\begin{array}{c} 
Q_1 \\
Q_2 \\
Q_3 \\
Q_4 \\
Q_5
\end{array}\right)^\pm
\label{eq:renpc_sub}
\end{equation}
In other words, first the lattice subtraction is performed, followed by the
$\chi RS$ renormalization. The above mixing pattern is abbreviated, in matrix
form, as
\bea
\hat Q^\pm &=& Z_\chi^\pm \tilde Q^\pm \nn \\
\tilde Q^\pm &=& [I+\Delta^\pm]Q^\pm
\eea
where $I$ is the $5 \times 5$ unit matrix and the subscript $\chi$ stands
for $\chi RS$ subtractions
\footnote{Attention is drawn to the notation adopted from now on in this paper:
whereas the elements of matrices $\ct{Z}^\pm$ and $\Delta^\pm$ are denoted, in
standard fashion, as $\ct{Z}_{ij}^\pm$ and $\Delta_{ij}^\pm$, those of matrix
$Z^\pm_\chi$ are denoted by $Z^\pm_{ij}$ (i.e. the subscript $\chi$ is dropped
for notational economy). The elements of matrix $Z^\pm$, appearing in
eq.~(\ref{eq:mrenpc}), will never be used from now on.}.
The renormalization of the parity
conserving sector is given by $Z^\pm = Z_\chi^\pm [I+\Delta^\pm]$.
Note that, using continuous chiral transformations,  in the hypothetical $\chi RS$, 
it is easy to show that $\Delta^\pm=0$
and $\ct{Z}_{ij}=Z_{ij}$. 

\section{NP renormalization: the lattice RI renormalization scheme}
\label{sec:mixing}

In this section, we outline the strategy used in the 
determination of the renormalization constants, and discuss the 
appropriate renormalization conditions.
According to the NP method proposed in \cite{NP}, the renormalization 
conditions are imposed in momentum space on the projected amputated 
Green's functions.  In this work we will always consider Green's functions 
where all external quark legs have the same momentum $p$.  This choice is,
of course,
not unique but is the simplest way to regulate the infrared divergences.   
An important point is that the renormalization conditions need to be imposed
at large Euclidean $p^2$.
In the determination of the mixing coefficients this condition enables us
to neglect soft or spontaneous chiral symmetry breaking effects which
are not induced by the Wilson term and cannot be computed.
As demonstrated in sec.~\ref{sec:rcwi}, only at large external momenta is
the NP method, described below, equivalent to the WI one.
The large $p^2$ condition is also necessary in the evaluation of the
overall renormalization constants, since the standard procedure of
obtaining physical amplitudes from renormalized matrix elements requires
perturbative matching with the Wilson coefficients in the continuum,
at fixed gauge coupling. We will describe in subsect.~\ref{subsec:range} 
an alternative procedure, for which this condition can be relaxed.

The renormalization method presented in this section is an extension of the 
one introduced in \cite{DS=2} for the renormalization of the
$O^{\Delta S =2}$ four-fermion operator. As in ref.~\cite{DS=2}, we will
define projected-amputated Green functions, on which suitable renormalization
conditions will be imposed. With respect to ref.~\cite{DS=2}, the novelties
consist in using a complete operator basis for the parity-conserving sector
and in extending the method to the parity-violating operators.

\subsection{Amputated Green functions and their projectors}
\label{subsec:ampproj}

We first give some general definitions. Since the non-perturbative
renormalization conditions are to be imposed  on  quark states, we shall need
the general expression of the four-point Green's functions of the operators 
$O^{\pm}_{\Gamma^{(1)}\Gamma^{(2)}}$.
Denoting by $x_1,x_3$ and $x_2,x_4$  the coordinates of the  
outgoing and incoming quarks respectively, we define 
the connected one-particle-irreducible Green's functions as
\begin{equation}
G^\pm_{\Gamma^{(1)}\Gamma^{(2)}}(x_1,x_2,x_3,x_4)=
\<\psi_1(x_1)\bar\psi_2(x_2) O^\pm_{\Gamma^{(1)}\Gamma^{(2)}}(0)
\psi_3(x_3)\bar\psi_4(x_4)\>\, ,  
\label{eq:G_Gamma(x)}
\end{equation}
where $\<\cdots\>$ denotes the vacuum expectation value, 
i.e.\  the average over the gauge-field configurations.
The generic four-fermion operator $O_{\Gamma^{(1)}\Gamma^{(2)}}$, placed at the
origin, is written as
\bea
\label{eq:O_Gamma}
&& O^\pm_{\Gamma^{(1)}\Gamma^{(2)}}(0) = \\
&& \frac{1}{2}\left[
\bar\psi_1(0)\Gamma^{(1)m}\psi_2(0) \bar\psi_3(0) \Gamma^{(2)m} \psi_4(0) 
\pm \bar\psi_1(0)\Gamma^{(1)m}\psi_4(0) \bar\psi_3(0)\Gamma^{(2)m}\psi_2(0) 
\right]\, .
\nonumber
\eea
Given the complexity of the notation, we shall elaborate on this formula.
In the case of parity
conserving operators, the two Dirac matrices are equal; i.e.\  
we are dealing with the $O^\pm_{\Gamma\Gamma}$ of eq.~(\ref{eq:qgamgam}).
The case of two different Dirac matrices
$\Gamma^{(1)} \neq \Gamma^{(2)}$ applies to the parity violating operators of
eqs.~(\ref{vaav24}),(\ref{sppsm24}) and (\ref{spps24}), for which, 
for example,
$\Gamma^{(1)} = V$ and $\Gamma^{(2)} = A$. Note that
$G^\pm_{\Gamma^{(1)}\Gamma^{(2)}}$ depends implicitly on the four colour
and Dirac indices carried by the external fermion fields. These
will be shown explicitly from now on. Following the conventions of
Appendix \ref{app:not}, upper-case Latin superscripts ($A,B,\dots,R,S,\dots$)
will denote colour indices of the fundamental SU(3) representation.
Greek lower-case subscripts ($\alpha,\beta,\dots,\rho,\sigma,\dots$) 
will denote spinor indices. The letters $m$ and $n$ are reserved
for the Lorentz indices of the Dirac matrices. Repeated Lorentz indices
are summed according to the convention adopted in Appendix \ref{app:not}.

The Fourier transform of the non-amputated Green's function 
(\ref{eq:G_Gamma(x)}), at equal external momenta $p$, has the form
\begin{equation}
G^\pm_{\Gamma^{(1)}\Gamma^{(2)}}(p)^{ABCD}_{\alpha\beta\gamma\delta}
  =\frac{1}{2}\left[
   \<\Gamma^{(1)m}(p)^{AB}_{\alpha\beta} 
\Gamma^{(2)m}(p)^{CD}_{\gamma\delta}\>
\mp\<\Gamma^{(1)m}(p)^{AD}_{\alpha\delta}
\Gamma^{(2)m}(p)^{CB}_{\gamma\beta}\>
\right]\, ,
\label{eq:G_Gamma(p)}
\end{equation}
where
\begin{equation}
\Gamma^{(i)m}(p)^{AB}_{\alpha\beta}
=S(p|0)^{AR}_{\alpha\rho}\Gamma^{(i)m}_{\rho\sigma}
 [\gamma_5S(p|0)^{\dag}\gamma_5]^{RB}_{\sigma\beta}, \qquad (i = 1, 2)
\label{eq:gammaa}
\end{equation}
and $S(p|0)$ is defined by
\begin{equation}
S(p|0)=\int d^4x S(x|0) e^{-ip\cdot x}.
\end{equation}
$S(x|0)$ is the inverse of the lattice Dirac operator; 
i.e. it is the ``quark propagator'' computed 
on a single gauge-field configuration and is therefore not translationally
invariant (cf. section 4 of ref.~\cite{NP}). It satisfies the relation 
\begin{equation}
S(x|0)=\gamma_5 S^{\dag}(0|x)\gamma_5.
\end{equation} 
The amputated Green's function is obtained from eq.\ 
(\ref{eq:G_Gamma(p)}) 
\begin{equation}
\Lambda^{\pm}_{\Gamma^{(1)}\Gamma^{(2)}}(p)^{RSR'S'}_{\rho\sigma\rho'\sigma'} 
=S^{-1}(p)^{RA}_{\rho\alpha}S^{-1}(p)^{R'C}_{\rho'\gamma}
 G^{\pm}_{\Gamma^{(1)}\Gamma^{(2)}}(p)^{ABCD}_{\alpha\beta\gamma\delta}
  S^{-1}(p)^{BS}_{\beta\sigma}S^{-1}(p)^{DS'}_{\delta\sigma'} 
\label{eq:Lambda_Gamma(p)}
\end{equation}
where $S(p)=\<S(p|0)\>$ is the Fourier transform of the translationally 
invariant quark propagator, 
i.e.\   the Fourier transform of $S(x|0)$, averaged over the
gauge-field configurations.

The amputated Green's function of eq.~(\ref{eq:Lambda_Gamma(p)}) is a 
high-rank tensor, from which a
more manageable function of the external momenta $p$ can be obtained by 
projecting over all the possible Dirac structures.  Let us introduce a 
generic Dirac projector 
\begin{equation}
\Pj_{\hat\Gamma^{(1)}\hat\Gamma^{(2)}}\equiv
 (\hat\Gamma^{(1)n}\otimes \hat\Gamma^{(2)n}),
\end{equation}
Its application on the amputated Green's functions is defined as
\footnote{ 
Other possible choices are admissible; see \cite{BURAS}. }:
\begin{equation}
\Tr\Pj_{\hat\Gamma^{(1)}\hat\Gamma^{(2)}} 
\Lambda^\pm_{\Gamma^{(1)}\Gamma^{(2)}} (p) =
(\hat\Gamma^{(1)n}_{\sigma\rho}\otimes \hat\Gamma^{(2)n}_{\sigma'\rho'})
 \Lambda^{\pm}_{\Gamma^{(1)}\Gamma^{(2)}}(p)^{RRR'R'}_{\rho\sigma\rho'\sigma'},
\label{eq:proj_def_1}
\end{equation}
The trace is taken over spin and colour.
The projectors $\Pj_{\hat\Gamma^{(1)}\hat\Gamma^{(2)}}$ can be worked 
out analytically for the tree-level amputated Green's function
\begin{equation}
\Lambda^{\pm (0)}_{\Gamma^{(1)}
\Gamma^{(2)}}(p)^{RSR'S'}_{\rho\sigma\rho'\sigma'}
=\frac{1}{2}\left[ \delta^{RS}\delta^{R'S'}
(\Gamma^{(1)m}_{\rho\sigma}\otimes\Gamma^{(2)m}_{\rho'\sigma'})
\mp \delta^{RS'}\delta^{R'S}
(\Gamma^{(1)m}_{\rho\sigma'}\otimes\Gamma^{(2)m}_{\rho'\sigma})\right]
\label{eq:Lambda_+_0}
\end{equation}
The superscript $(0)$ denotes tree-level.
From eqs.~(\ref{eq:proj_def_1}) and (\ref{eq:Lambda_+_0}) one finds
\begin{eqnarray}
\Tr\Pj_{\hat\Gamma^{(1)}\hat\Gamma^{(2)}}
\Lambda^{\pm (0)}_{\Gamma^{(1)} \Gamma^{(2)}}(p) 
&=&(\hat\Gamma^{(1)n}_{\sigma\rho}\otimes \hat\Gamma^{(2)n}_{\sigma'\rho'})
 \Lambda^{\pm (0)}_{\Gamma^{(1)}\Gamma^{(2)}}(p)^{RRR'R'}_{\rho\sigma\rho'
\sigma'} \nn \\
&=&\frac{1}{2} [ N_c^2 (\Tr \hat\Gamma^{(1)n}\Gamma^{(1)m})
(\Tr \hat\Gamma^{(2)n}\Gamma^{(2)m})
\nn \\
& & \mp N_c(\Tr \hat\Gamma^{(1)n}\Gamma^{(1)m}
\hat\Gamma^{(2)n}\Gamma^{(2)m}) ], 
\label{eq:pjpv0}
\end{eqnarray}
These formulae can now be used in the specific cases of interest, namely the
renormalization of the various parity violating and parity
conserving operators (i.e. $Q^\pm_k$ of eq.~(\ref{eq:qq})
and $\ct{Q}^\pm_k$ of eq.~(\ref{eq:hh}) with
$k=1,\dots,5$). We will see shortly that, for the renormalization
conditions we will impose on the quark correlation functions
of the four-fermion operators, it is convenient to use projectors that
obey the following orthogonality conditions:
\begin{eqnarray}
\label{eq:orth}
\Tr\Pj^\pm_i \Lambda^{\pm (0)}_k = \delta_{ik} \qquad (i,k = 1,\dots,5) \\
\Tr \Pjpv ^\pm_i {\ctLambda}^{\pm (0)}_k =
\delta_{ik} \qquad (i,k = 1,\dots,5) \nonumber
\end{eqnarray}
where $\Lambda^{(0) \pm}_k$ and $\ctLambda^{(0) \pm}_k$ are the tree-level
amputated Green functions of operators $Q^\pm_k$ and ${\ct{Q}}^\pm_k$
respectively ($k=1,\dots,5$),
and $\Pj^\pm_k$, $\Pjpv^\pm_k$ their corresponding projectors ($k=1,\dots,5$).
In the following we will denote as $\Lambda^\pm_k$ and $\ctLambda^\pm_k$
the corresponding Green functions in the interacting case.
With the aid of eq.~(\ref{eq:pjpv0}) we find for the parity conserving case
\begin{eqnarray}
&& \Pj^{\pm}_1 \equiv + \frac{1}{64N_c(N_c\pm 1)}(\Pj_{VV}+\Pj_{AA}) \nn \\
&& \Pj^{\pm}_2 \equiv + \frac{1}{64(N_c^2-1)}(\Pj_{VV}-\Pj_{AA}) 
               \pm \frac{1}{32N_c(N_c^2-1)}(\Pj_{SS}-\Pj_{PP}) \nn \\
&& \Pj^{\pm}_3 \equiv \pm \frac{1}{32N_c(N_c^2-1)}(\Pj_{VV}-\Pj_{AA})
               +  \frac{1}{16(N_c^2-1)}(\Pj_{SS}-\Pj_{PP})  
\label{eq:projqpc} \\
&& \Pj^{\pm}_4 \equiv + \frac{(2N_c\pm 1)}{32N_c(N_c^2-1)}(\Pj_{SS}+\Pj_{PP})
                \mp \frac{1}{32N_c(N_c^2-1)}\Pj_{TT}  \nn \\
&& \Pj^{\pm}_5 \equiv \mp \frac{1}{32N_c(N_c^2-1)}(\Pj_{SS}+\Pj_{PP})
               + \ \frac{(2N_c\mp 1)}{96N_c(N_c^2-1)}\Pj_{TT}  \nn
\end{eqnarray}
whereas for the parity violating one
\begin{eqnarray}
&& \Pjpv^{\pm}_1 \equiv - \frac{1}{64N_c(N_c\pm 1)}(\Pj_{VA}+\Pj_{AV}) \nn \\
&& \Pjpv^{\pm}_2 \equiv - \frac{1}{64(N_c^2-1)}(\Pj_{VA}-\Pj_{AV})
                 \mp \frac{1}{32N_c(N_c^2-1)}(\Pj_{SP}-\Pj_{PS})  \nn \\
&& \Pjpv^{\pm}_3 \equiv \mp \frac{1}{32N_c(N_c^2-1)}(\Pj_{VA}-\Pj_{AV})
                    -  \frac{1}{16(N_c^2-1)}(\Pj_{SP}-\Pj_{PS})  
\label{eq:projqpv} \\
&& \Pjpv^{\pm}_4 \equiv + \frac{(2N_c\pm 1)}{32N_c(N_c^2-1)}(\Pj_{SP}+\Pj_{PS})
                 \mp \frac{1}{32N_c(N_c^2-1)}\Pj_{T\tilde T}  \nn \\
&& \Pjpv^{\pm}_5 \equiv \mp \frac{1}{32N_c(N_c^2-1)}(\Pj_{SP}+\Pj_{PS})
                   + \frac{(2N_c\mp 1)}{96N_c(N_c^2-1)}\Pj_{T\tilde T} \nn 
\end{eqnarray}

\subsection{Renormalization conditions}
\label{sub:ren_pv_q1}

We now define the renormalization procedure of the operators of interest.
The general principle is to impose ``suitable" renormalization conditions,
which are satisfied by the renormalized (projected amputated) Green functions
$\hat \Lambda^\pm$ and $\hat{\ctLambda} ^\pm$ at a fixed scale $\mu$ in the
deep Euclidean region. The renormalization condition is arbitrary.
A simple choice is to impose that the fully interacting
$\hat \Lambda^\pm$ (and $\hat{\ctLambda} ^\pm$), at a given scale $\mu$, are
equal to their tree level values written in eq.~(\ref{eq:pjpv0}); see also 
eqs.~(\ref{eq:orth}).

We will use matrix notation for simplicity: the amputated-projected Green
functions $\Lambda^\pm$ and $\ctLambda ^\pm$ denote $1 \times 5$ row
vectors, whereas the projectors $\Pj^\pm$ and $\Pjpv^\pm$ denote
$5 \times 1$ column vectors. In this notation, eqs.~(\ref{eq:orth}) become
\bea
\Pj^\pm \cdot \Lambda^{\pm (0)} = I \nn \\
\Pjpv^\pm \cdot \ctLambda ^{\pm (0)} = I
\label{eq:orthm}
\eea
with $I$ the $5\times 5$ unit matrix. For the parity-violating case,
the renormalized amputated Green's function is given by the row vector
\be
\hat{\ctLambda}^\pm(p/\mu,g^2) =
Z_{\psi}^{-2}(a\mu,g^2_0) \ctLambda ^\pm(ap,g^2_0) \ct{Z}^\pm (a\mu,g^2_0)^T
\label{eq:lhat}
\ee
where $Z_\psi$ is the quark field renormalization constant
and $\ct{Z}^\pm$ is the $5\times 5$ renormalization matrix (the superscript
$T$ stands for transpose). Recall that $\ct{Z}^\pm$ is the block diagonal
matrix of eq.~(\ref{eq:renpv}). In the above expression we denote by
$g_0^2 \equiv g_0^2(a)$ the bare coupling and by $g^2 \equiv g^2(\mu)$ the
renormalized one. We express the bare Green
function in terms of a ``dynamics" matrix $\ct{D}$ from which the tree-level
amputated Green's function is factored out:
\begin{equation} 
\ctLambda ^\pm = \ctLambda ^{\pm (0)} \ct{D}^\pm 
\label{eq:dyn23}
\end{equation}
Note that since the matrix $\ct{D}$ determines the ``dynamics" of the
bare operators, it can only mix tree-level operators with the same
discrete symmetries. Thus, it is also a block diagonal matrix with the same
block structure as $\ct{Z}^\pm$.
From eqs. (\ref{eq:orthm}) and (\ref{eq:dyn23}) the elements of the matrix
$\ct{D}^\pm$ are expressed in terms of the amplitudes $\ctLambda ^{\pm}$:
\begin{equation}
\ct{D}^\pm = \Pjpv ^\pm \cdot \ctLambda ^\pm
\label{eq:dmat23}
\end{equation}
We compute $\ctLambda ^\pm$ non-perturbatively, at fixed coupling $g_0^2$
and in a given gauge, over a configuration ensemble. We opt for the Landau
gauge. From $\ctLambda ^\pm$ and eq.~(\ref{eq:dmat23}), we obtain $\ct{D}^\pm$.

The renormalization conditions
\be
\label{eq:Z(mu23)}
\Pjpv ^{\pm} \cdot \hat{\ctLambda} ^\pm (p/\mu,g^2) 
\Bigg \vert _{p^2=\mu^2} = I
\label{eq:rc2345_pv}
\ee
determine the mixing matrix $\ct{Z}^\pm$. This is easily seen by
combining eqs.~(\ref{eq:orthm}), (\ref{eq:lhat}), (\ref{eq:dmat23})
and (\ref{eq:rc2345_pv}) to obtain
\be
Z_{\psi}^{\pm -2} \ct{D}^\pm \ct{Z} ^{\pm T} = I
\label{eq:rcpvoz}
\ee
from which we obtain $\ct{Z}^\pm$ in terms of the known quantities
$Z_{\psi}^{\pm 2}$ and $\ct{D}^\pm$:
\be
\ct{Z} ^\pm = Z_{\psi}^{\pm 2} \left[ \ct{D}^{\pm \, T}\right]^{-1}
\label{eq:rcpvoz2}
\ee 
Note that since both $\ct{D}^\pm$ and $\ct{Z}^\pm$ have the same block diagonal
structure, this involves inverting at most $2 \times 2$ matrices.

The quark field renormalization $Z_\psi$ is also determined non-perturbatively
from the numerical simulation. A definition which respects WI's (cf.~sec. 4
of ref.~\cite{NP} for details) is
\be
Z_\psi=-i\frac{1}{12}\Tr \gamma_\mu\frac{\partial S(p)^{-1}}{\partial p_\mu}
\Bigg \vert _{p^2=\mu^2},
\ee
Instead, we have implemented
\begin{equation}
Z_\psi(\mu a)=\left.-i\frac{1}{12}\frac{\sum_{\mu=1,4}\gamma_\mu \sin(p_\mu a)S(pa\
)^{-1}}{4\sum_{\mu=1,4}\gamma_\mu \sin^2(p_\mu a)}\right|_{p^2=\mu^2}
\label{eq:Z_psi}
\end{equation}
in order to avoid derivatives w.r.t. discrete variables.

The computation of the parity-conserving matrix $Z^\pm$ follows similar
lines. Recall that this matrix can be cast in the form $Z_\chi^\pm [ I +
\Delta^\pm ]$, with $Z_\chi^\pm$ the block diagonal matrix of 
eq.~(\ref{eq:renor_subt}) and $\Delta^\pm$ the sparse matrix of 
eq.~(\ref{eq:renpc_sub}). We find it convenient to obtain $Z_\chi^\pm$ and
$\Delta^\pm$ separately. As we will show in the next section,
$Z_\chi^\pm$ depends on the renormalization scale $a\mu$, whereas $\Delta^\pm$
does not. 
More importantly, as we will also show in sect. \ref{sec:rcwi}, the
independence of $\Delta^\pm$ from $a\mu$ is due to the fact that the lattice
subtractions can also be determined from WI's, in the spirit of
ref.~\cite{jlqcd_lat96}. Thus, our results on $\Delta^\pm$, obtained in the RI
scheme with the NP method, can in principle be compared to those obtained with
WI's.

For the parity-conserving case, the fully interacting renormalized amputated
Green's function is given by the row $1 \times 5$ vector
\be
\hat \Lambda ^\pm(p/\mu,g^2) =
Z_{\psi}^{-2}(a\mu,g^2_0) \Lambda ^\pm(ap,g^2_0) [I + \Delta^{\pm}(g_0^2)^T ]
Z_\chi^\pm (a\mu,g^2_0)^T
\label{eq:lhat2}
\ee
We express the bare Green's function in terms of a ``dynamics" matrix
$D^\pm$:
\begin{equation}
\Lambda ^\pm = \Lambda ^{\pm (0)} D^\pm 
\label{eq:dyn23pc}
\end{equation}
From eqs.~(\ref{eq:orthm}) and (\ref{eq:dyn23pc}) the elements of the matrix
$D^\pm$ are expressed in terms of the amplitudes $\Lambda ^{\pm}$:
\begin{equation}
D^\pm = \Pj ^\pm \cdot \Lambda ^\pm
\label{eq:dmat23pc}
\end{equation}
Therefore $D^\pm$ can be computed non-perturbatively by numerical simulation,
at fixed bare coupling $g_0^2$, over a configuration ensemble and in the
Landau gauge.
Once it is computed, we determine the mixing matrix $\Delta^\pm$ and the
renormalization matrix $Z_\chi^\pm$ from the renormalization conditions
\be
\Pj ^{\pm} \cdot \hat{\Lambda} ^\pm (p/\mu,g^2) 
\Bigg \vert _{p^2=\mu^2} = I
\label{eq:rc2345_pc}
\ee
To see this explicitly, we combine eqs.(\ref{eq:orthm}), (\ref{eq:lhat2}),
(\ref{eq:dyn23pc}) and (\ref{eq:dmat23pc}), in order to express
eq.(\ref{eq:rc2345_pc}) as follows:
\be
Z_{\psi}^{-2} D^\pm [I + \Delta^{\pm \,\,^T} ] Z_\chi^{\pm \,\,T} = I
\label{eq:lhat2mat}
\ee
We then proceed in two steps. Let us first rewrite the above expression as
\be
D^\pm_{kl} + \sum_{j=1}^5 D^\pm_{kj} \Delta^\pm_{lj} =
Z_{\psi}^2 \left[ Z_\chi^{\pm \,\,{-1}} \right]_{lk}
\ee
(in matrix-component notation). We then consider the special case, with indices
$k,l$ chosen so that for a given fixed value $k$, the index $l$ is allowed to
run over the corresponding ``lattice subtracted'' values. For example, if
we fix $k=1$, $l$ runs over the range $l=2,3,4,5$; if we fix $k=2$ or $k=3$,
$l$ runs over the range $l=1,4,5$ etc. With these choices of $k,l$, and given
the structure of matrix $\Delta^\pm$ (c.f. eq.(\ref{eq:renpc_sub})), the 
summed index $j$ also runs over the same interval as $l$ (all other
contributions involve zero matrix elements of $\Delta^\pm$). Moreover, the
r.h.s. vanishes for these combinations of $k,l$, due to the block-diagonal
structure of the matrix $Z_\chi^\pm$ (c.f. eq.(\ref{eq:renor_subt})).
Defining the column vector $c^\pm_l = D^\pm_{kl}$ for fixed $k$ and $l$ running
as detailed above, we obtain the equation
\be
\Delta^\pm_{lj} c^\pm_j = - c^\pm_l
\ee
This is a linear inhomogenous system, which can be solved for
$\Delta^\pm_{lj}$.
Thus, the mixing matrix is determined, and the subtracted correlation function
can be constructed. We can now proceed to the second step, which is the
determination of the renormalization matrix $Z_\chi^\pm$. It can be obtained
exactly like in the parity violating case; cf. eqs.~(\ref{eq:rcpvoz})
and (\ref{eq:rcpvoz2}).

\subsection{Range of validity of the RI renormalization scheme}
\label{subsec:range}
Having completed our discussion on the determination of the renormalization
constants and mixing coefficients of four-fermion operators,
using the NP method of ref.~\cite{NP}, we now summarize the renormalization
conditions. For the parity violating operators we have
\bea
\label{eq:rirs1}
&& Z_\psi^{-2} \Tr \Pjpv ^{\pm} \cdot
\ctLambda ^\pm \ct{Z} ^{\pm \,\,T} \Bigg \vert _{p^2=\mu^2} = I
\eea
For the parity conserving operators we have
\bea
Z_\psi^{-2} \Tr \Pj^{\pm} \cdot
\Lambda^\pm Z^{\pm \,\, T} \Bigg \vert _{p^2=\mu^2} =
Z_\psi^{-2} \Tr \Pj^{\pm} \cdot
\tilde \Lambda^\pm Z^{\pm \,\, T}_\chi \Bigg \vert _{p^2=\mu^2} &=&
\nonumber \\
Z_\psi^{-2} \Tr \Pj^{\pm} \cdot \Lambda^\pm 
[ I + \Delta^{\pm \,\, T} ] Z^{\pm \,\, T}_\chi
\Bigg \vert _{p^2=\mu^2} &=& I
\label{eq:rirs3}
\eea
with $Z_\psi$ determined from eq.~(\ref{eq:Z_psi}). In the last equation,
$\tilde \Lambda^\pm$ denotes the amputated Green's function for the 
lattice-subtracted parity conserving operators.
These conditions constitute a renormalization scheme,
called the Regularization Independent (RI) scheme \cite{NP,CIUCHINI2}
(also known as the MOM scheme).  The name RI is chosen so as
to distinguish it from the $\MSbar$ scheme, adopted in perturbation
theory, which depends on the detailed choice of dimensional
regularization ('t Hooft-Veltman, Dimensional Reduction, etc.).
In the RI scheme, the renormalized operator depends on the gauge and on the
momenta of the external states \cite{NP,DS=2}. Unlike the $\MSbar$, the RI
scheme does not depend on the regularization.
Since the renormalized matrix element must be multiplied by the Wilson 
coefficient $C_W(\mu)$ (cf.~eq.~(\ref{eq:ope})) the latter must also be 
calculated in the same gauge and with the same external states in order to
obtain a renormalization group invariant result. The Wilson coefficient is
known in next-to-leading order continuum PT \cite{CIUCHINI2}; thus, its 
matching with the renormalized matrix element is accurate to that order.
Clearly, for the perturbative calculation of $C_W(\mu)$ to be reliable,
we must ensure that $\mu \gg \Lambda_{\QCD}$.

There is also another reason for
which this condition must be imposed: the RI conditions are such that the
renormalized operators transform according to irreducible representations
of the chiral algebra. This is achieved by imposing that the projections of
the renormalized operators on the ``wrong'' chiral structures vanish (cf.
eq.(\ref{eq:rirs3})). But this not true when
there are other causes of chiral symmetry breaking, either due to the explicit
presence of mass terms or due to spontaneous symmetry breaking in the chiral
limit. Such effects, present in QCD, would modify the r.h.s. of the
renormalization conditions (\ref{eq:rirs1}) and (\ref{eq:rirs3}) by form
factors proportional to the quark mass $m$, the chiral condensate
$\langle \bar \psi \psi \rangle$ and the inverse scale $\mu^{-1}$.
These terms die off in the large momenta region, $\mu \gg \Lambda_{\QCD}$.

On the other hand,
the non-perturbative renormalization of the four-fermion operator in the RI
scheme involves computations of bare matrix elements at finite lattice
cut-off. In order to have good control of the discretization errors, 
we must also ensure that $\mu \ll \ct{O}(a^{-1})$. Thus, as already pointed out
in \cite{NP}, the RI scheme is applicable at couplings $g_0^2$ and lattice
sizes for which there exists a window of $\mu$ values satisfying
\be
\Lambda_{\QCD} \ll \mu \ll \ct{O}(a^{-1})
\label{eq:wind}
\ee

The discussion leading to the necessity of the above window of $\mu$ values
is based on the assumption that the operator renormalization is performed at
fixed UV cutoff (i.e. the gauge coupling is fixed in the numerical
computation). One can relax the  bound of eq.~(\ref{eq:wind}) by
performing a sequence of computations at bare couplings $g_0(a)$, $g_0(sa)$,
$g_0(s^2 a)$, $\cdots$ corresponding to increasingly coarse lattice spacings
(i.e. the scaling factor $s > 1$; typically $s=2$). This method of lowering
the renormalization
scale $\mu$ in a controlled NP way is reminiscent of the renormalization
procedure of ref.~\cite{sharperome}, which in turn was inspired by
refs.~\cite{lusch}. We  outline it here in schematic fashion; wave function
renormalization, operator subtractions etc. will be omitted for the sake of
clarity. Realistically, with present day  resources, this procedure can  be
iterated only two or three times.

We first select a small coupling constant $g_0^2$, so that we are safely in
the perturbative region (i.e. $g_0^2$ is  smaller than current values
in  standard QCD simulations). For such small lattice spacings, 
the window of eq.~(\ref{eq:wind})
is easily satisfied by both momentum scales $\mu$ and $\mu/s$ used in the
following. We now chose a  lattice of $N$ lattice sites in
each direction ($N$ is only limited by current computational capabilities),
in such a way that  $a\mu \ll 1$, in order to avoid 
(UV) lattice artifacts, and at the same time large enough to have negligible
finite size effects; i.e. $a\mu/s \gg 1/N$. 
With small $g_0^2 \sim 0.7$, and $N\sim 32$--$48$,
this lattice is adequate for simulations of quark correlation functions
deep in the perturbative region, but cannot accomodate hadronic quantities,
due to the smallness of its physical volume. Once the RI scheme has been
implemented at the scale $\mu$, the renormalized correlation function of
a given operator is known at any  scale $p$ satisfying eq.~(\ref{eq:wind}):
\be
\hat \Lambda \left( \frac{p}{\mu}, g^2(\mu) \right) =
Z \left ( a\mu,g_0^2 \left (a \right ) \right) \Lambda \left (ap,
g_0^2 \left (a \right ) \right )
\label{eq:sc1}
\ee
In particular, we define
\be \hat \Lambda_1 \equiv \hat \Lambda \left( \frac{1}{s}, g^2(\mu) \right)
\label{eq:lll} \ee
i.e. the renormalized vertex at a scale $p=\mu/s$. We now  increase the
lattice spacing to  $a^\prime = s a$. This is done in practice by tuning the
bare coupling $g_0^2$, on a bigger (coarser) lattice.  For simplicity, since
we assumed that finite size effects are negligible both at $\mu$ and $\mu/s$,
we keep $N$ fixed. On the coarser lattice we compute the
bare correlation function at momentum $p^\prime = \mu/s$ (i.e. for
$a^\prime p^\prime = a\mu$). The renormalization constant at the scale $\mu$
(for cutoff $a^\prime$) can then be obtained from the equation
\be
\hat \Lambda_1  =
Z \left ( a^\prime \mu, g_0^2 \left (a^\prime \right ) \right)
\Lambda \left (a^\prime p^\prime,  g_0^2 \left ( a^\prime \right ) \right )
\label{eq:sc2}
\ee
since the l.h.s. of this expression is known from the calculation on the finer
lattice (eq.~(\ref{eq:lll})). With the renormalization constant
$Z \left ( a^\prime \mu, g_0^2 \left (a^\prime \right ) \right)$
thus obtained on the coarse lattice, we now compute the bare correlation
function at momentum $p^{\prime\prime} = \mu/s^2$ (i.e. 
$a^\prime p^{\prime\prime} = a\mu/s$). For this momentum we then have
\be
\hat \Lambda \left( \frac{1}{s^2}, g^2(\mu) \right) =
Z \left ( a^\prime \mu, g_0^2 \left (a^\prime \right ) \right) \Lambda
\left (a^\prime p^{\prime \prime},  g_0^2 \left ( a^\prime \right ) \right )
\label{eq:sc2b}
\ee
Thus the renormalized correlation function is now known at a lower momentum
scale $1/s^2$:
\be
\Lambda_2 \equiv \hat \Lambda \left( \frac{1}{s^2}, g^2(\mu) \right)
\ee
One can now repeat the cycle: going to a coarser lattice
$a^{\prime \prime} = s^2 a$, we compute the bare correlation
function at momentum $p^{\prime \prime} = \mu/s^2$ (i.e.
$a^{\prime \prime} p^{\prime \prime} = a \mu$). The renormalization constant
at scale $\mu$ is then obtained by solving
\be
\Lambda_2 = Z \left ( a^{\prime\prime} \mu, g_0^2 \left(a^{\prime\prime}
\right ) \right)
\Lambda \left (a^{\prime\prime} p^{\prime\prime},
g_0^2 \left( a^{\prime\prime} \right ) \right )
\ee
for $ Z \left ( a^{\prime\prime} \mu, g_0^2 \left(a^{\prime\prime}
\right ) \right)$. The bare correlation function is then computed at momentum
$p^{\prime \prime \prime} = \mu/s^3$ (i.e. $a^{\prime \prime}
p^{\prime \prime \prime} = a\mu/s$) and renormalized by $Z$ at scale $\mu$,
giving the renormalized correlation function at scale $1/s^3$:
\be
\hat \Lambda \left( \frac{1}{s^3}, g^2(\mu) \right) =
Z \left ( a^{\prime \prime} \mu, g_0^2 \left (a^{\prime \prime} \right ) 
\right) \Lambda \left (a^{\prime \prime} p^{\prime \prime \prime}, 
g_0^2 \left ( a^{\prime \prime} \right ) \right )
\ee
At the end of the day, by keeping $N$ fixed, we end up with a lattice
coarse enough to contain the hadrons, and an operator, renormalized 
non-perturbatively at a low scale $\mu/s^n$.

\section{Mixing coefficients from WI's}
\label{sec:rcwi}

In \cite{jlqcd_lat96} an alternative approach to the NP renormalization
method has been applied. This is based on lattice WI's on quark
states, in the spirit of \cite{BOCHICCHIO} and \cite{lucgui,ZETA_A}.
The WI renormalization has been implemented in \cite{jlqcd_lat96}
for the non-perturbative evaluation of the mixing
renormalization constants of the $\Delta S=2$ matrix element of
$K^0 - \bar K^0$ oscillations. In our language, this is the renormalization
of the parity conserving operator $Q_1^+$. The WI's were derived in momentum
space, the various chiral structures were  projected out (just like in
\cite{DS=2} and the present work) and they were solved for the mixing lattice
coefficients $\Delta^+_{1k}$ (for $k=2,\dots,5)$. The multiplicative
renormalization constant $Z^+_{11}$ was obtained using the NP method of
\cite {NP,DS=2}.
In principle, both methods are equivalent (under certain conditions, which
will be discussed below) and suffer from the same
sources of systematic error. In practice, one may give more stable results
than the other. A systematic comparison of the two methods is an interesting
problem; for the $\Delta S = 2$ case it has been carried out in
ref.~\cite{jlqcd_lat96}.

Here we wish to extend this method to the complete basis of operators.
We will therefore write down the WI's which can be used for the extraction
of all mixing coefficients (i.e. all the elements of the matrix $\Delta^\pm$
of eq.~(\ref{eq:renpc_sub})). We will also demonstrate explicitly that the WI
method for the determination of the lattice mixing coefficients is equivalent
to the RI method. This equivalence is true, up to discretization errors,
provided  the renormalization scale $\mu$ of the RI scheme is large
(cf. eq.~(\ref{eq:wind})) and that we are working in the chiral limit
or with a mass independent scheme. The reasons for
requiring a large $\mu$ for the RI scheme have been explained in
subsect.~\ref{subsec:range}. We now present briefly the reasons for which
the implementation of the small $m$ limit for the WI scheme is required.

The WI holds for operators with the correct chiral properties, i.e.
multiplicatively renormalizable operators transforming according to a well
defined representation of the chiral algebra. By imposing
its validity on the renormalized operators one can fix the mixing coefficients
of the operators which stem from the chiral violation due to the Wilson
term. This, however, is only true in the chiral limit, where chiral symmetry
breaking is only due to the presence of the Wilson term in the lattice action.
Upon introducing a mass term (soft symmetry breaking), the continuum
WI's can only be recovered by the simultaneous redefinition of the $T$-product
and the operator. This procedure is ambiguous, since a change in the definition
of the $T$-product can be compensated by a redefinition of the operator.
Thus, away
from the chiral limit, the WI's do not uniquely define the operator (and,
consequently, its mixing coefficients). This ambiguity is harmless if we
can apply low-energy theorems of Current Algebra (small quark mass).
This point has been discussed in ref.~\cite{BOCHICCHIO}, and
more recently  in \cite{testa}.

We will explicitly demonstrate in this section that the WI method and the RI
renormalization scheme are equivalent methods for the determination of the
lattice mixing coefficients, under the conditions discussed above. Moreover,
we will show that the $\chi RS$ renormalization matrices $\ct{Z}$ and $Z_\chi$
cannot be determined from WI's.
Only finite ratios of $\chi RS$ renormalization constants of opposite parity
are fixed by the WI's. Thus the RI scheme (or some other renormalization
method) is necessary for their determination, even when the WI method is used
in order to obtain the lattice mixing coefficients.

\subsection{Ward Identities on quark states}
\label{subsec:wiqs}

The validity of the above statements has been shown, in a general and elegant
way, in ref.~\cite{testa}. Our aim is much more specific: we will obtain
useful WI's which can be used in practice for the determination of the
mixing coefficients of all four-fermion operators of interest. The above
general statements will consequently be proved in the context of these
specific WI's, which can be implemented in numerical simulation. In what
follows, for reasons of uniformity of presentation, we will continue to work
with operators which have four distinct flavours (unlike
ref.~\cite{jlqcd_lat96}, where the flavour group used was $SU(2)$ and the
$\Delta S=2$ operator considered carried the strange and down physical
flavours). The most economical way of obtaining useful WIs for the four
flavour operators defined in eqs.~(\ref{eq:qq}) and (\ref{eq:hh}) consists
in the following trick. We consider an ``embedding'' of our operators (with
four distinct flavours) in a theory with five flavours. In other words, our
flavour symmetry group is $SU(5)_L \otimes SU(5)_R$
\footnote{In the end of this subsection, we will comment on the relevance
of our results to the physical case of three light quark flavours.}.
A suitable field variation
of the fifth flavour will then yield WIs concerning the operators of
eqs.~(\ref{eq:qq}) and (\ref{eq:hh}).
Finally, since one of our aims is to propose WI's which can be used in
simulations, we will be working with a small finite quark mass. It
is to be understood, throughout the rest of this section, that the chiral
limit is to be taken in the end.

We now establish our notation. For the purposes of this Section 
$\pm$ superscripts will be dropped in what follows from operators
and correlation functions (for example, from the
$Q^\pm$'s, $G^\pm$'s and $\Lambda^\pm$'s). We will be using
Green functions which are expectation values of the following operators:
\bea
\nonumber
G_k (x_0;x_1,x_2,x_3,x_4)^{ABCD}_{\alpha\beta\gamma\delta} &=&
\psi_1(x_1)^A_\alpha \bar \psi_2(x_2)^B_\beta
Q_k(x_0) \psi_3(x_3)^C_\gamma \bar \psi_4(x_4)^D_\delta \nonumber \\
\ct{G}_k (x_0;x_1,x_2,x_3,x_4)^{ABCD}_{\alpha\beta\gamma\delta} &=&
\psi_1(x_1)^A_\alpha \bar \psi_2(x_2)^B_\beta \ct{Q}_k(x_0)
\psi_3(x_3)^C_\gamma \bar \psi_4(x_4)^D_\delta
\label{eq:defops}
\eea
for $k=1,\dots,5$. The operators $Q_k$ and $\ct{Q}_k$ on the r.h.s. are
those of eqs.~(\ref{eq:qq}) and (\ref{eq:hh}); i.e. they are defined in terms
of fermion fields with flavours $1,\cdots,4$. Their vacuum expectation
values are the four-quark correlation functions of eq.(\ref{eq:G_Gamma(x)}),
which are the starting point of the $RI$ renormalization procedure. We will
also be using five other operators (and their four-quark correlation functions)
which are defined as follows:
\bea
\breve{\ct{G}_k} (x_0;x_1,x_2,x_3,x_4)^{ABCD}_{\alpha\beta\gamma\delta}
&=& \psi_1(x_1)^A_\alpha \bar \psi_2(x_2)^B_\beta \breve \ct{Q}_k(x_0)
\psi_3(x_3)^C_\gamma \bar \psi_4(x_4)^D_\delta
\nonumber \\
\ct{G}_k^5 (x_0;x_1,x_2,x_3,x_4)^{ABCD}_{\alpha\beta\gamma\delta} &=&\left[
\gamma_5 \psi_5(x_1) \right]^A_\alpha \bar \psi_2(x_2)^B_\beta \breve
\ct{Q}_k(x_0) \psi_3(x_3)^C_\gamma \bar \psi_4(x_4)^D_\delta 
\label{eq:defops2}
\eea
where
\bea
\breve \ct{Q}^\pm_k = \left( \bar \psi_5 \Gamma^{(1)} \psi_2 \right)
\left( \bar \psi_3 \Gamma^{(2)} \psi_4 \right) \pm
\left( \bar \psi_5 \Gamma^{(1)} \psi_4 \right)
\left( \bar \psi_3 \Gamma^{(2)} \psi_2 \right)
\eea
and $k=1, \cdots ,5$ stands for the combination of Dirac matrices
corresponding to the definitions of eq.~(\ref{eq:hh}). In other words,
$\breve \ct{Q}^\pm_k$ is obtained from $\ct{Q}^\pm_k$ by substituting
$\bar \psi_1$ by $\bar \psi_5$. The operators $\ct{G}_k^5$ are obtained from
$\ct{G}_k$, with the substitution $(\psi_1, \bar \psi_1) \rightarrow
(\gamma_5 \psi_5, \bar \psi_5)$.
The colour and spin indices ($A,B,\dots$ and $\alpha, \beta,\dots$) will be
dropped from now on for notational simplicity.
We will also need the axial current and pseudoscalar density
\bea
A_\mu(x) &=& \bar \psi_1(x) \gamma_\mu \gamma_5 \psi_5(x)
\nonumber \\
P(x) &=& \bar \psi_1(x) \gamma_5 \psi_5(x)
\eea

We perform infinitesimal axial transformations on quark fields in terms
of the $SU(5)$ raising operator
\begin{equation}
\begin{array}{c} 
\lambda =
\end{array}
\left(\begin{array}{rrrrr}
0 & 0 & 0 & 0 & 1 \\
0 & 0 & 0 & 0 & 0 \\
0 & 0 & 0 & 0 & 0 \\
0 & 0 & 0 & 0 & 0 \\
0 & 0 & 0 & 0 & 0
\end{array}\right)
\label{eq:lambdaI}
\end{equation}
which induces transformations between the first and fifth flavours only:
\bea
\psi_1 \rightarrow \psi_1 + i \delta \alpha \gamma_5 \psi_5
\nonumber \\
\bar \psi_5 \rightarrow \bar \psi_5 + i \delta \alpha \bar \psi_1 \gamma_5
\label{eq:axq}
\eea
Correspondingly, operators $\breve \ct{Q}$ transform into their
$Q$ counterparts:
\bea
\breve \ct{Q}_1 &\rightarrow& \breve \ct{Q}_1 - i \delta \alpha Q_1
\nonumber \\
\breve \ct{Q}_2 &\rightarrow& \breve \ct{Q}_2 + i \delta \alpha Q_2
\nonumber \\
\breve \ct{Q}_3 &\rightarrow& \breve \ct{Q}_3 + i \delta \alpha Q_3
\label{eq:QI} \\
\breve \ct{Q}_4 &\rightarrow& \breve \ct{Q}_4 + i \delta \alpha Q_4
\nonumber \\
\breve \ct{Q}_5 &\rightarrow& \breve \ct{Q}_5 + i \delta \alpha Q_5
\nonumber
\eea
Whenever convenient, matrix notation will be used, as in
subsect.~\ref{sub:ren_pv_q1}. We define the $1 \times 5$ row vectors
\bea
\breve \ct{G} &=& (\breve \ct{G}_1, \breve \ct{G}_2, \breve \ct{G}_3,
\breve \ct{G}_4, \breve \ct{G}_5)
\nonumber \\
G &=& (-G_1, G_2, G_3, G_4, G_5) 
\label{eq:vects}\\
\ct{G}^5 &=& (\ct{G}^5_1, \ct{G}^5_2, \ct{G}^5_3, \ct{G}^5_4, \ct{G}^5_5)
\nonumber
\eea
Note the minus sign in the element $G_1$ of vector $G$; it has been introduced
in its definition in order to take into account the minus sign of the axial
variation of $\breve \ct{Q}_1$; c.f. eq.(\ref{eq:QI}).

The lattice WI arising from this axial variation of the $1 \times 5$
row vector $\langle {\breve \ct{G}} \rangle$ is
\bea
\frac{\delta}{\delta \alpha(x)} \langle \breve \ct{G}(x_0;x_1,x_2,x_3,x_4)
\rangle = 0 \Leftrightarrow \nonumber \\
\langle \frac{\delta}{\delta \alpha(x)} \breve \ct{G}(x_0;x_1,x_2,x_3,x_4)
\rangle =
\langle \breve \ct{G}(x_0;x_1,x_2,x_3,x_4) \frac{\delta S}{\delta \alpha(x)}
\rangle
\eea
where all points $x_0,x_1,x_2,x_3$ and $x_4$ are kept separate. The variation
of the operator $\breve \ct{G}$ receives a contribution $\langle G \rangle$
from the variation of $\breve \ct{Q}$ and a contribution from the variation
of the fermion field $\psi_1$. Integrating over $x$ yields
\bea
\langle G \rangle + \langle \ct{G}^5 \rangle =
- \int d^4 x \langle \breve \ct{G}
\left[ \nabla^\mu_x A_\mu(x) - 2 m_0 P(x) - X_5(x) \right] \rangle
\label{eq:wi1}
\eea
In the chiral limit, the term $2 m_0 P(x)$ on the r.h.s. vanishes, whereas
the integrated total divergence of the axial current gives a non-zero
surface term, due to the presence of Goldstone bosons. If we are not in
the chiral limit, the $2 m_0 P(x)$ term is present, but the surface term
from the current divergence vanishes upon integration. We will be considering
the latter case, in order to mimic what is happening in the simulations
(i.e. first we compute at small non-zero quark mass and then extrapolate
to the chiral limit). 
The operator $X_5$ arises from the variation of the chiral symmetry breaking
Wilson term in the action. As shown in ref.~\cite{BOCHICCHIO} (see also
\cite{crluvl} for a detailed discussion) it mixes, under
renormalization with $\nabla^\mu A_\mu$ and $P$. This mixing, determined
by the requirement that on-shell matrix elements of the subtracted $X_5$
vanish in the continuum, generates a finite renormalization of the axial
current and a power subtraction of the quark mass. Thus, following
ref.~\cite{BOCHICCHIO}, in the above WI we will trade-off $X_5$ for the
renormalized expression $[\nabla^\mu \hat A_\mu - 2 \hat m \hat P -
\overline X_5]$, where $\overline X_5$ is the subtracted $X_5$.
What is of interest to us is that, besides the above renormalizations, the 
$\overline X_5$ insertion in the above correlation function also generates
contact terms. They are found by looking at the flavour content and discrete
symmetries of the specific correlation function. We find (up to Schwinger
terms which will vanish under the integral):
\bea
\langle \overline X_5(x) \breve \ct{G}_k(x_0;x_1,x_2,x_3,x_4) \rangle &=&
\langle G_j \rangle \left( \delta_{jk} - R_{jk} \right) \delta(x -x_0)
\nonumber \\
&-& \langle G_j \rangle \Delta_{ij} R_{ik}
\delta(x-x_0)
\label{eq:xct}
\eea
where repeated indices are summed; $i,j = 1,\cdots,5$.
The notation for the various coefficients has been chosen with some
foresight: The $5 \times 5$ matrix $R_{jk}$ is a block diagonal matrix of
the form of $Z_\chi$ of eq.~(\ref{eq:renor_subt}). It will eventually
turn out that $R = Z_\chi^T (\ct{Z}^{-1})^T$; i.e. $R$ is the finite ratio of
the parity-conserving to parity-violating, logarithmically diverging,
renormalization matrices of $\chi RS$ type. The $5 \times 5$ matrix
$\Delta_{ij}$ is a sparse matrix of the form of eq.~(\ref{eq:renpc_sub});
it will eventually turn out to be the matrix defined in
eq.~(\ref{eq:renpc_sub}). Finally note that the product $[\Delta^T R]_{jk}$
is a sparse matrix of the form of $\Delta$. In conclusion, we have separated
the contact terms on the r.h.s. of eq.~(\ref{eq:xct}) into a first term which
mixes the $\langle G_j \rangle$'s as in the $\chi RS$ scheme and a second term
which generates lattice mixing. Note that in the above expression, we have
only included the contact terms generated by the proximity of
$\overline X_5(x)$ to the operators $\ct{Q}(x_0)$. In principle there are
also contact terms arising from the proximity of $\overline X_5(x)$ to the
quark fields of the correlation $\ct{G}$ at points $x_1,\cdots,x_4$. However,
as shown in ref.~\cite{testa}, these terms vanish in the continuum limit. The
reason is that the insertion of the operator $\overline X_5$ with fundamental
fields (once both this operator and the quark fields are renormalized) is
proportional to the lattice spacing.

We now combine eqs.~(\ref{eq:wi1}) and (\ref{eq:xct}), Fourier-transform
the WI (with all external momenta set equal to $p$) and amputate the
resulting correlation functions. We denote by $\Lambda(p)$ and
$\ctLambda^5 (p)$ the momentum-space amputated Green functions of 
$\langle G \rangle$ and
$\langle \ct{G}^5 \rangle$. The Fourier-transformed $\breve \ct{G}$ is
denoted by $\breve \ct{G} (p)$. The resulting momentum-space amputated WI is
\bea
\Lambda (p) \left[ I + \Delta^T \right] R  =
- \ctLambda^5 (p) +
2 \hat m \int d^4 x \langle \breve \ct{G} (p)
\hat P(x) \rangle \prod_1^4 \langle S^{-1}(p) \rangle
\label{eq:wi1mom}
\eea
We require that the above WI, up to quark field renormalization,
be the one valid in the continuum limit for renormalized correlation functions
(operators) and that it be identical to the corresponding nominal WI
(which is just the tree-level version of the above equation). This implies
that $\Delta$ is indeed the matrix of lattice subtractions. Moreover,
$R$ can indeed be identified with the ratio of the renormalization matrices
$Z_\chi^T (\ct{Z}^{-1})^T$, since this combination would renormalize both
sides of the above equation (NB: the renormalization of $\breve \ct{G}(p)$
is identical to that of $\ct{G}(p)$, since the two operators only differ by
a relabelling of $\psi_5$ as $\psi_1$).

We now show how the WI fixes both the lattice mixing matrix $\Delta$ and the
matrix ratio $R$. We consider the first column of the above WI; its l.h.s.
concerns the operator $Q_1$ and its lattice mixings. By projecting this WI
with five projectors of eqs.~(\ref{eq:projqpc}) and (\ref{eq:projqpv}) we
construct a linear inhomogenous system of five equations w.r.t. the quantities
$R_{11}$ and $[\Delta^T R]_{k1} = R_{11} \Delta_{1k}$, with $k=2,\cdots,5$.
By solving this system, the lattice mixing of $Q_1$ and the ratio of the
multiplicative renormalization constants $R_{11} = Z_{11}/\ct{Z}_{11}$ of
$Q_1$ and $\breve \ct{Q}_1$ (which is just $\ct{Q}_1$ with a flavour
relabelling) can be determined. Similarly, a system of ten equations 
(using 10 projectors on the second and third columns of the WI)
determines the lattice mixing of $Q_2$ and $Q_3$ (i.e. the $\Delta_{jk}$'s
with $j=2,3$ and $k=1,4,5$) and the elements $R_{ik}$ with $i,k=2,3$. The case
of $Q_4$ and $Q_5$ is identical to that of $Q_2$ and $Q_3$ (but involves the
fourth and fifth rows of the WI).

So far we have shown how WIs can determine the lattice mixing coefficients
and the finite ratio of the $\chi RS$ renormalization matrices. We will
now show that these quantities are compatible to the ones obtained from
the RI renormalization conditions, in the
large momentum limit. We project both sides of eq.~(\ref{eq:wi1mom}) with the
$5 \times 1$ column vector $\Pj$, defined as
\be
\Pj = (-\Pj_1, \Pj_2, \Pj_3, \Pj_4, \Pj_5)^T
\ee
(the minus sign in $\Pj_1$ corresponds to that of $-G_1$ in
eq.~(\ref{eq:vects}). For large $p$, the last term on the r.h.s.
vanishes. This is because the explicit $m_0$ factor implies that the
integrated term has one less dimension that the others so that, at large
momenta it vanishes faster by one power of $p$ (see ref.~{\cite{testa}).
Moreover, in this limit the inverse quark propagator behaves as
$S^{-1}(p) = i \Sigma_1 \gamma_\mu p_\mu$ (with $\Sigma_1$ a scalar form
factor and up to ${\cal O}(a)$ terms). This means that in
this limit it anticommutes with $\gamma_5$. Combining this result with the
definition of $\Pj_k$ we easily deduce that
\bea
\Tr \Pj_k \ctLambda _j^5 &=& - \Tr \Pjpv _k \ctLambda_j
\label{eq:pwi1}
\eea
(with $k,j=1,\cdots,5$). Thus, the WI becomes
\be
\Tr \Pj \cdot \Lambda \left[ I + \Delta^T \right] Z_\chi^T
= \Tr \Pjpv \cdot \ctLambda \ct{Z}^T
\ee
This WI, given the RI renormalization condition for the parity violating
operator
\be
\Tr \Pjpv \cdot \ctLambda \ct{Z}^T = I
\ee
implies the RI renormalization condition of eq.~(\ref{eq:rirs3}) 
for the parity conserving operators $Q$.

We now comment briefly on the relation of these results to the more realistic
case of three light flavours. The physical operators $O^-$ always mix with
others of lower dimension. This introduces further complications which are
beyond the scope of the present work. For the physical operators $O^+$,
there are penguin-type contributions which, however, cancel. Thus, the 
resulting WIs are identical to the ones obtained here with five flavours.

\subsection{Scale dependence of the renormalization constants}
\label{subsec:scale}

We will now specify which of these renormalization constants are
divergent quantities in the UV limit and which are finite. First
we give a general discussion of
the functional dependence of the $Z$'s (and $\ct{Z}$'s)
on the coupling, mass, renormalization point and cutoff. We are interested
in the physics of light quark masses, so we assume $\mu \gg m$.
The $Z$'s are dimensionless quantities; thus, in principle, they could 
have the functional dependence $Z(g_0^2,a\mu,am,m/\mu)$, where
$m$ stands for the degenerate quark mass and $\mu$ is the renormalization
scale (i.e. it stands for a generic choice of the four external quark momenta).
It must be chosen so as to regularize all IR divergences, including those
arising in the chiral limit. For the four-fermion operators we are considering
in this work, it suffices to simply take equal momenta; i.e. $p_i^2 = \mu^2$
($i=1,\dots,4$ for the four external legs; $\mu$ is space-like.).
Any regular dependence on $am$ and $a\mu$ should drop out in the continuum limit
$a \rightarrow 0$, and is therefore treated in simulations
as a systematic error due to the finiteness of the cutoff. These
errors are, say, $\ct{O}(am)$, $\ct{O}(a\mu)$  for the Wilson action and
$\ct{O}(amg_0^2)$, $\ct{O}(a\mu g_0^2)$
for the tree level improved Clover action, used in the present paper.
Moreover, the $Z$'s cannot
be singular in $am$ (e.g. have $\ln(am)$ terms in the small mass limit
$m \ll \mu$) because they would
diverge not only in the continuum limit, but also in the chiral limit
($m \to 0$). Similarly, regular terms in $m/\mu$ are neglected (light masses),
whereas singular ones must be absent (existence of chiral limit).
In conclusion, for light quark masses, the functional dependence of the
$Z$'s is in general of the form $Z(g_0^2,a\mu)$. 
\par
Power counting suggest that their dependence on $a\mu$ is singular
(logarithmic). However, this is only true of the $\chi RS$ renormalization
constants $Z_\chi$ and $\ct{Z}$. The lattice mixing coefficients
$\Delta$ and the ``ratio" $\ct{Z}^{-1}Z_\chi$ can be determined
from a system of equations (the projected WI's discussed above).
This implies that  they can be expressed in terms of bare lattice correlation
functions, which do not depend on the renormalization scale. Thus their
functional dependence is of the form $Z(g_0^2)$; i.e. they are finite.
\par
Finally, we can lift the mass degeneracy, introducing small mass differences
$\delta m$. This could in principle introduce dependences like powers
of $a \delta m$ and $\ln(m_i/m_j)$ in the $Z$'s.
However, as pointed out in \cite{BERNARD2}, none of these
survive: The regular terms (e.g. powers of $a\delta m$) vanish
in the continuum limit $a \rightarrow 0$; i.e. they are part of the usual
$O(am)$ and $O(g_0^2 am)$ discretization errors. Singular (power or
logarithmic) dependence
is not allowed by the requirement that the chiral limit of the theory 
($m_i \rightarrow 0$) exist and it be well defined. Thus, lifting the mass 
degeneracy does not spoil the renormalization pattern discussed above.

\subsection{Identities between renormalization constants}
\label{subsec:idsrc}

In this section we derive useful identities which relate some parity conserving
renormalization constants of $\ct{S}=1$ operators to renormalization
constants of their $\ct{S}=-1$ counterparts. These identities are formally
exact, but are only approximately satisfied in practical computations.
Thus, they are useful tests of the reliability of our results.

In order to derive them, we first need to show that, once the operator basis
has been renormalized using the RI scheme, any other basis of
renormalized operators (formed by linear combinations of the original
renormalized operators) also satisfies the RI conditions. This is very
straightforward. In the first basis, the RI scheme consists in the
following properties of the projected amputated Green functions $\Lambda$:
\bea
\Tr \Pj \Lambda^{(0)} & = & I \nonumber \\
\hat \Lambda & \equiv & Z \Lambda
\label{eq:basis1_ri}
\\
\Tr \Pj \hat \Lambda & = & \Tr \Pj (Z \Lambda) = I \nonumber 
\eea
The above are valid at the renormalization scale $p^2 = \mu^2$. This
shorthand notation should be clear to the reader.
Now we can define a new basis of operators, obtained by a rotation $R$ of
the original basis. This implies rotated projected amputated
Green functions $\Lambda'$ and rotated projectors ${\Pj}'$, which must
satisfy:
\bea
\Lambda' & \equiv & R \Lambda \nonumber \\
\Tr {\Pj}' \Lambda^{(0)'} & = & I
\eea
Trivially, the last equation is satisfied, provided that
\be
{\Pj}' = \Pj R^{-1}
\label{eq:ppp}
\ee
It is straightforward to combine eqs.(\ref{eq:basis1_ri})--(\ref{eq:ppp})
to show that also $\hat \Lambda'$ satisfies the RI renormalization
condition $\Tr {\Pj}' \hat \Lambda' = I$. The implication of this property is
that if we use the RI scheme in order to renormalize separately the operators
$O_{[VV+AA]}$ and $O^F_{[VV+AA]}$,  for example, the renormalized
operator $\hat Q_{[VV+AA]} + \hat Q_{[VV+AA]}^F$ satisfies the RI scheme; in
other words it
is the renormalized operator $\hat Q_1^+$. In this section, we will make
use of this property. We also note that in this section we will
exclusively work with the lattice subtracted parity conserving operators
$\tilde Q_k$ and the parity violating operators
$\ct{Q}_k$. As discussed in subsecs. \ref{subsec:chiralsub} and
\ref{subsec:wiqs}, these operators have good chiral properties
(in other words, they can be thought of as operators in the $\chi RS$).

We consider the operator $O_{[VA-AV]}$, which only
mixes with the operator $O_{[SP-PS]}^F$. Thus, we renormalize it in the RI
scheme to obtain
\be
\hat O_{[VA-AV]} =  z_{22} O_{[VA-AV]} - z_{23} O_{[SP-PS]}^F
\label{eq:q2subtrq3}
\ee
The same renormalization pattern is obeyed by operator $O^F_{[VA-AV]}$ (it only
mixes with $O_{[SP-PS]}$) since it only involves a relabelling of flavours:
\be
\hat O^F_{[VA-AV]} = z_{22} O^F_{[VA-AV]} - z_{23} O_{[SP-PS]}
\label{eq:q2subtrq3F}
\ee
The last two equations can be combined into
\be
\hat \ct{Q}_2^\pm = \hat O_{[VA-AV]} \pm \hat O^F_{[VA-AV]}
= z_{22} \ct{Q}_2^\pm \pm z_{23} \ct{Q}_3^\pm
\ee
which implies that
\bea
\ct{Z}^+_{22} &=& \ct{Z}^-_{22} = z_{22} \nn \\
\ct{Z}^+_{23} &=& - \ct{Z}^-_{23} = z_{23}
\label{eq:z2232pm}
\eea
Similar expressions can be derived for the renormalization constants
of operators $\ct{Q}_3^\pm$ and the parity conserving ones in the
$\chi RS$ scheme. In practical
simulations, we expect that these identities are well satisfied in the
parity violating case. For the parity conserving one (which also involves
lattice subtractions) the agreement should only be approximate.

We note the above proof rests onto two crucial properties of the operators
$O_{[VA-AV]}$ and $O_{[SP-PS]}^F$ concerned: (i) they mix with each other
only; (ii) they transform into each other
under Fierz transformations in Dirac space (cf. eq.(\ref{eq:fautopc})).
These two properties determine the mixing pattern of eqs.~(\ref{eq:q2subtrq3})
and (\ref{eq:q2subtrq3F}). This proof cannot be extended to operators
$\ct{Q}_4^\pm$ and $\ct{Q}_5^\pm$. The second requirement is satisfied by the
pair of
operators $O_{[SP+PS]}$ and $O_{[SP+PS]}^F - O_{[T\tilde T]}^F$; see
eq.~(\ref{eq:fautopc}). However, the first requirement is not satisfied,
since operator $O_{[SP+PS]}$ mixes not only with
$O_{[SP+PS]}^F - O_{[T\tilde T]}^F$, but also with
$O_{[SP+PS]} - O_{[T\tilde T]}$ (cf. Table \ref{tab:g12}). Thus, no property
analogous to eq.~(\ref{eq:z2232pm}) can be found for the operators
$\ct{Q}^\pm_4$ and $\ct{Q}^\pm_5$ (or a linear combination of them).

\section{Numerical results}
\label{sec:res}

In order to test the feasibility of these ideas, we have computed the
renormalization constants, in the RI scheme, of the complete basis of
operators $Q_k$ and $\ct{Q}_k$, for $k = 1,\cdots,5$. This was done in
the quenched approximation, for the Wilson and tree-level improved Clover
action. We have performed simulations at three values of the
gauge coupling, namely $\beta=6/g_0^2=6.0,~6.2,~6.4$. Our results were
obtained at finite quark masses and extrapolated to the chiral limit $\kappa_c$.
The specific values of all lattice parameters used in these simulations
can be found in tab.~\ref{tab:params}.
\setlength{\tabcolsep}{.16pc}
\begin{table}[htb]
\begin{center}
\begin{tabular}{||l||c|c|c|c|c|c||}
\hline\hline
$\beta$ & 6.0  & 6.0    & 6.2 & 6.2    & 6.4 & 6.4  \\
\hline
Action   & C   & W & C  & W & C  & W \\
\# Confs & 100  & 100    & 180 & 100    & 60  & 60     \\
Volume   &$16^3\times 32$ & $16^3\times 32$ & $16^3\times 32$ & $16^3\times 32$
         &$24^3\times 32$ & $24^3\times 32$ \\
\hline
$\kappa$ &0.1425& 0.1530 & 0.14144 & 0.1510 & 0.1400 & 0.1488 \\
         &0.1432& 0.1540 & 0.14184 & 0.1515 & 0.1403 & 0.1492 \\
         &0.1440& 0.1550 & 0.14224 & 0.1520 & 0.1406 & 0.1496 \\
         &      &        & 0.14264 & 0.1526 & 0.1409 & 0.1500 \\
\hline
$\kappa_c$ & 0.14551 & 0.15683 & 0.14319 & 0.15337 & 0.14143 & 0.15058 \\
\hline\hline
$a^{-1}$&2.16(4)& 2.26(5)& 2.70(10) & 3.00(9) & 4.00(20)& 4.10(20)\\  
\hline\hline
\end{tabular}
\end{center}
\vspace{0.3truecm}
\caption{Parameters of the runs used for the NP calculation of 
the renormalization constants.  We also give the critical value of the
hopping parameter $\kappa_c$ and the inverse lattice spacing $a^{-1}$,
as quoted in ref.~\cite{calib}.}
\label{tab:params}
\end{table}
We also give in the table the value of the inverse lattice spacing $a^{-1}$,
obtained in ref.~\cite{calib} on the same dataset. All statistical errors have
been estimated with the jacknife method, decimating 10 configurations at a
time.

As has been explained in sec.~\ref{sec:mixing}, the NP method is based on the
computation of quark Green's functions; thus gauge-fixing has
to be implemented. We have worked in the lattice Landau gauge, defined by
minimizing the functional
\begin{equation}
\Tr~\left[\sum_{\mu=1}^4\left( U_{\mu}(x)+U^{\dag}_{\mu}(x)\right) \right].
\end{equation}
Possible effects from Gribov copies have been ignored. However, 
in analogy to the study of the effect of Gribov ambiguities on the
renormalization of two-quark operators of ref.~\cite{Paciello}, we expect them
to be small.

We now present detailed results for one representative case, namely the
renormalization constants of the Clover action at $\beta = 6.2$, as a function
of the renormalization scale in lattice units. All results are extrapolated in
the chiral limit. In Figs.~\ref{fig:z4fpWILL} we plot the renormalization
constants of operators $Q^+_1$ and $\ct{Q}^+_1$. We see that both $\chi RS$
renormalization constants $Z_{11}^+$ and $\ct{Z}_{11}^+$ are scale dependent,
whereas the mixing coefficients $\Delta_{1k}$ ($k=2,\dots 5$)
become more stable with increasing $\mu$. This is what we expect
from the discussion of sect.~\ref{sec:rcwi}.
In figs.~\ref{fig:z4fpWILR} and \ref{fig:z4fpWILRMix} we show similar
results for the renormalization constants of the operators $Q^+_k$
and $\ct{Q}_k^+$, for $k = 2,3$, whereas in figs.~\ref{fig:z4fpWITT}
and \ref{fig:z4fpWITTMix} we show those for $k=4,5$.
Most $\chi$RS renormalization constants of the
matrix $Z_\chi^+$  display a marked $\mu$-dependence due to a
non-zero  anomalous dimension,
whereas the lattice subtraction coefficients of the matrix $\Delta^+$
are roughly scale-independent in the window $a\mu \in [1,2]$.
Compared to the others, the renormalization constants $\Delta^+_{24}$,
$\Delta^+_{34}$ and $\Delta^+_{43}$ have a more pronounced
variation with the scale $\mu$; nevertheless they are reasonably flat in the
same window.

In order to give a more quantitative flair of our results, we present
the renormalization matrices, for all actions and couplings, in
Appendix~\ref{app:numerical}. Only results at the renormalization scale
$\mu \simeq 2$ GeV are presented. Note that the identities derived in
subsec.~\ref{subsec:idsrc} for the $\chi RS$ renormalization constants of
operators $\ct{Q}_2^\pm$ and $\ct{Q}_3^\pm$ (c.f. eq.(\ref{eq:z2232pm})) are
well satisfied. For the parity conserving operators $Q_2^\pm$ and $Q_3^\pm$
similar identities appear to be approximately true.
At the scale $\mu \simeq 2$GeV, these results can be directly
used in the computation of the matrix elements of the corresponding
operators. A first implementation of these results in the calculation of
various $B$-parameters can be found in refs.~\cite{b78910,Bsusy2}.

The validity of our results should be confirmed independently
from WIs, computed with the same actions and at the same couplings.
Work in this direction is in progress.

\section*{Acknowledgements}
We warmly thank E.~Franco, L.~Giusti, G.C.~Rossi and M.~Testa for many
stimulating and thorough discussions. A careful reading of the manuscript by
G.C.~Rossi is also gratefully acknowledged. V.G. acknowledges the partial
support by CICYT under grant number AEN-96/1718.
\appendix
\section{Notation and conventions}
\label{app:not}

The 16 Euclidean Dirac $4\times 4$ 
matrices, which form a complete basis, are denoted by
\begin{equation}
\Gamma
=\{\id,\gamma_{\mu}, \sigma_{\mu\nu}, \gamma_{\mu}\gamma_5,\gamma_5\}
\equiv \{S,V,T,A,P\},
\end{equation}
where
\bea
&& \gamma_5 \equiv -\frac{1}{4!} \epsilon_{\mu\nu\rho\lambda}
 \gamma_\mu \gamma_\nu \gamma_\rho \gamma_\lambda =
 - \gamma_0 \gamma_1 \gamma_2 \gamma_3
\nonumber \\
&& \sigma_{\mu\nu}=\frac{1}{2}[\gamma_{\mu},\gamma_{\nu}]
\eea
with $\epsilon_{\mu\nu\rho\lambda}$ the completely antisymmetric
rank- four pseudotensor with $\epsilon_{0123} = +1$.
The Euclidean Dirac matrices satisfy the following properties
\begin{equation}
\{\gamma_{\mu},\gamma_{\nu}\}=2\delta_{\mu\nu},\qquad
\gamma_{\mu}^{\dag}=\gamma_{\mu},\qquad \gamma_5^{\dag}=\gamma_5.
\end{equation}
We also define the dual ``sigma" matrix
\be
\tilde \sigma_{\mu\nu} \equiv \frac{1}{2} \epsilon_{\mu\nu\rho\lambda}
\sigma_{\rho\lambda} = \gamma_5 \sigma_{\mu\nu} \equiv \tilde T
\ee
The helicity projectors are, as usual,
\[
\gamma^L_\mu \equiv L_\mu \equiv \gamma_\mu (1-\gamma_5),\qquad
\gamma^R_\mu \equiv R_\mu \equiv \gamma_\mu (1+\gamma_5).
\]
and
\[
L \equiv (1-\gamma_5),\qquad
R \equiv (1+\gamma_5).
\]
Repeated $\Gamma$ matrices imply summation of their Lorentz indices (if any); 
for example $VV \equiv \sum_\mu \gamma_\mu \otimes \gamma_\mu$,
$VA \equiv \sum_\mu \gamma_\mu \otimes \gamma_\mu \gamma_5$, etc. Note however, 
that $TT = \sigma_{\mu \nu} \otimes \sigma_{\mu \nu}$ and
$T \tilde T = \sigma_{\mu \nu} \otimes \tilde \sigma_{\mu \nu}$ means
summation over the 6 independent $\sigma_{\mu \nu}$ matrices (e.g. summation
over $\mu$ and $\nu$ with $\mu > \nu$).

The colour group is $SU(N_c)$ with $N_c=3$. The Gell- Mann group generators
are denoted by $t^a, a=1,\dots,N^2_c-1$. Fermion fields $\psi^{A\alpha}$
carry spinor and colour indices.
Latin uppercase letters denote colour indices in the fundamental representation
($A,B,\dots = 1,\dots, N_c$), whereas Greek lowercase letters stand for 
Dirac spinor indices ($\alpha,\beta,\dots,\rho,\sigma,\dots = 1,\dots,4$). 
The letters $m$ and $n$ are reserved for Lorentz indices running on the set of
Dirac matrices (as in $\Gamma^{(1)m}\Gamma^{(2)m}$) according to the following
convention: if we are dealing with pairs of Dirac matrices such as $SS$, $SP$
etc., the index $m$ is absent. If we have pairs like $VV$, $VA$ etc, $m$
runs over the four values of $\gamma_\mu$.
In the case of the pairs $TT$ and $T\tilde T$, $m$
runs over the six independent values of the Dirac matrices
$\sigma_{\mu\nu}$ and $\tilde \sigma_{\mu\nu}$.

\section{Fierz transformations in Dirac space}
\label{app:Fierz}

In this Appendix we gather several useful formul\ae~concerning Fierz
transformations in Dirac space; colour indices are ignored.
We express the Fierz transformation of the Dirac indices of
a four fermion operator as follows
\begin{equation}
\bm{\Gamma} \otimes \bm{\Gamma} \equiv \Gamma_{\alpha\beta} \otimes
\Gamma_{\gamma\delta} \rightarrow \left[ {\bm{\Gamma}} \otimes
{\bm{\Gamma}} \right]^{F_D} \equiv \Gamma_{\alpha\delta} \otimes \Gamma_{\gamma\beta}
\end{equation}
The Euclidean Fierz transformed Dirac tensor products
$\left[ {\bm{\Gamma}} \otimes {\bm{\Gamma}} \right]^{F_D}$ can be 
reexpressed as a linear combination of the complete set of the
original tensor products $\bm{\Gamma} \otimes \bm{\Gamma}$, by exploiting
the completeness of the set of Dirac matrices.  One has 
\begin{equation}
\left(\begin{array}{c} 
\left[ {\bm{S}} \otimes \tilde{\bm{S}} \right]^{F_D} \\
\left[ {\bm{V}} \otimes \tilde{\bm{V}} \right]^{F_D} \\
\left[ {\bm{T}} \otimes \tilde{\bm{T}} \right]^{F_D} \\
\left[ {\bm{A}} \otimes \tilde{\bm{A}} \right]^{F_D} \\
\left[ {\bm{P}} \otimes \tilde{\bm{P}} \right]^{F_D} \\
\end{array}\right)
=-\frac{1}{4}
\left(\begin{array}{rrrrr}
1 &  1 & -1 & -1 & 1  \\
4 & -2 &  0 & -2 & -4  \\
-6 &  0 & -2 & 0 & -6  \\
-4 & -2 & 0 & -2 & 4  \\
 1 & -1 & -1 & 1 & 1  
\end{array}\right)
\left(\begin{array}{c} 
\bm{S} \otimes \bm{S} \\
\bm{V} \otimes \bm{V} \\
\bm{T} \otimes \bm{T} \\
\bm{A} \otimes \bm{A} \\
\bm{P} \otimes \bm{P} \\
\end{array}\right)
\label{eq:fidir}
\end{equation}
The overall minus sign is due to the anticommutativity of the Fermi fields.
Our Dirac $\Gamma$ matrices are normalized as
\begin{equation}
SS=1,\ VV=4,\ TT=-6,\  AA=-4,\ PP=1
\end{equation}
with the summation over Dirac indices understood. Recall that the sum in 
$TT$ runs only over the six independent $\sigma_{\mu \nu}$ matrices\footnote{
Sometimes $T$ and $A$ are defined with an imaginary $i$ prefactor
in order to have a positive 
normalization. This would imply a change of sign in the $T$ and $A$ rows and
columns of the $F$ matrix. This convention is not adopted here.}.

From the above equation, we can easily derive the following useful identities,
concerning parity conserving operators:
\begin{equation}
\begin{array}{lcl}
\left[ {\bm{V}} \otimes {\bm{V}} +
{\bm{A}} \otimes {\bm{A}} \right]^{F_D} &=&
\bm{V} \otimes \bm{V} + \bm{A} \otimes \bm{A} \nn \\
\left[ {\bm{V}} \otimes {\bm{V}} -
{\bm{A}} \otimes {\bm{A}} \right]^{F_D} &=&
- 2 \left [ \bm{S} \otimes \bm{S} - \bm{P} \otimes \bm{S} \right ] \nn \\
\left[ {\bm{S}} \otimes {\bm{S}} -
{\bm{P}} \otimes {\bm{P}} \right]^{F_D} &=&
-\dfrac{1}{2} \left [ \bm{V} \otimes \bm{V} - \bm{A} \otimes \bm{A} \right ]
\label{eq:fautopc} \\
\left[ {\bm{S}} \otimes {\bm{S}} +
{\bm{P}} \otimes {\bm{P}} \right]^{F_D} &=&
-\dfrac{1}{2} \left [ \bm{S} \otimes \bm{S} + \bm{P} \otimes \bm{P} 
- \bm{T} \otimes \bm{T} \right ] \nn \\
\left[ {\bm{S}} \otimes {\bm{S}} + {\bm{P}} \otimes {\bm{P}}
- \bm{T} \otimes \bm{T} \right]^{F_D} &=&
- 2 \left [ \bm{S} \otimes \bm{S} + \bm{P} \otimes \bm{P} 
\right ] \nn
\end{array}
\end{equation}
Moreover, by substituting $\psi_4 \rightarrow \gamma_5 \psi_4$ in the above
equations, we derive, for the parity violating operators $\ct{Q}^\pm_1$,
$\ct{Q}_2^\pm$ and $\ct{Q}_3^\pm$ of our basis:
\begin{equation}
\begin{array}{lcl}
\left[ {\bm{V}} \otimes {\bm{A}} + {\bm{A}} \otimes {\bm{V}} \right]^{F_D} &=&
\bm{V} \otimes \bm{A} + \bm{A} \otimes \bm{V} \nn \\
\left[ {\bm{V}} \otimes {\bm{A}} - {\bm{A}} \otimes {\bm{V}} \right]^{F_D} &=&
- 2 \left [ \bm{S} \otimes \bm{P} - \bm{P} \otimes \bm{S} \right ] \nn \\
\left[ {\bm{S}} \otimes {\bm{P}} - {\bm{P}} \otimes {\bm{S}} \right]^{F_D} &=&
-\dfrac{1}{2} \left [ \bm{V} \otimes \bm{A} - \bm{A} \otimes \bm{V} \right ]
\label{eq:fautopv} \\
\left[ {\bm{S}} \otimes {\bm{P}} +
{\bm{P}} \otimes {\bm{S}} \right]^{F_D} &=&
-\dfrac{1}{2} \left [ \bm{S} \otimes \bm{P} + \bm{P} \otimes \bm{S} 
- \bm{T} \otimes \bm{{\tilde T}} \right ] \nn \\
\left[ {\bm{S}} \otimes {\bm{P}} + {\bm{P}} \otimes {\bm{S}}
- \bm{T} \otimes \bm{{\tilde T}} \right]^{F_D} &=&
- 2 \left[ \bm{S} \otimes \bm{S} + \bm{P} \otimes \bm{P} 
\right] \nn
\end{array}
\end{equation}

Finally, from eq.~(\ref{eq:fidir}) we can easily derive the five linear
combinations of operators which are eigenstates of the transformation
with eigenvalues $\pm 1$; they are:
\bea
\left[ {\bm{S}} \otimes {\bm{S}} + {\bm{P}} \otimes 
{\bm{P}} + {\bm{T}} \otimes {\bm{T}} \right]^{F_D} &=& + \left[ 
\bm{S} \otimes \bm{S} + \bm{P} \otimes \bm{P} +\bm{T} \otimes \bm{T} \right]
\label{eq:fauto} \\
\left[ {\bm{V}}\otimes {\bm{V}}+ {\bm{A}} \otimes
{\bm{A}} \right]^{F_D} &=& + \left[ \bm{V} \otimes \bm{V} + \bm{A} \otimes
\bm{A} \right] \nn \\
\left[ 2({\bm{S}} \otimes {\bm{S}} - {\bm{P}} \otimes {\bm{P}}) - ({\bm{V}}
\otimes {\bm{V}} - {\bm{A}} \otimes {\bm{A}}) \right]^{F_D} &=&
+ \left[ 2(\bm{S} \otimes \bm{S} - \bm{P} \otimes \bm{P}) 
-(\bm{V} \otimes \bm{V} - \bm{A} \otimes \bm{A}) \right] \nn \\
\left[ 2({\bm{S}} \otimes {\bm{S}} - {\bm{P}} \otimes
{\bm{P}}) +({\bm{V}} \otimes {\bm{V}} - {\bm{A}}
\otimes {\bm{A}}) \right]^{F_D} &=&
- \left[ 2(\bm{S} \otimes \bm{S} - \bm{P} \otimes \bm{P}) 
+(\bm{V} \otimes \bm{V} - \bm{A} \otimes \bm{A}) \right] \nn \\
\left[ {\bm{S}} \otimes {\bm{S}} + {\bm{P}} \otimes
{\bm{P}} -\frac{1}{3} {\bm{T}} \otimes {\bm{T}} \right]^{F_D}
&=& - \left[ \bm{S} \otimes \bm{S} + \bm{P} \otimes \bm{P}
-\frac{1}{3}\bm{T} \otimes \bm{T} \right] \nn
\eea
Thus, in Dirac space, three linear combinations 
are Fierz eigenstates with eigenvalue $+1$
and two are Fierz eigenstates with eigenvalue $-1$.

\section{Numerical results}
\label{app:numerical}

In this appendix we present the renormalization matrices for both the Wilson
and Clover action at the three couplings considered. All results are in the
chiral limit. For each $\beta$ value, we show results at a scale $a\mu$ (in
lattice units) such that $\mu \simeq 2$GeV.

\subsection{Clover action, $\beta=6.0$, $\mu^2 a^2 =0.964$}

\[
{\cal Z}^- =
\left(
\begin{array}{rrrrr}
  0.90(4)& 0 & 0 & 0 & 0
\\
  0 & 0.97(3)& -0.45(1)& 0 & 0
\\
  0 & -0.017(3)& 0.36(1)& 0 & 0
\\
  0 & 0 & 0 & 0.27(1)&-0.024(6)
\\
  0 & 0 & 0 & 0.23(1)& 1.12(4)
\\
\end{array}
\right)
\]
\[
Z^-_\chi =
\left(
\begin{array}{rrrrr}
  0.92(4)& 0 & 0 & 0 & 0
\\
  0 & 0.99(3)& -0.42(1)& 0 & 0
\\
  0 & -0.019(4)& 0.47(1)& 0 & 0
\\
  0 & 0 & 0 & 0.36(2)&-0.024(6)
\\
  0 & 0 & 0 & 0.19(1)& 1.11(4)
\\
\end{array}
\right)
\]
\[
\Delta^- =
\left(
\begin{array}{rrrrr}
  0 &-0.15(1)&-0.02(1)&-0.15(2)&-0.09(1)
\\
 -0.23(1)& 0 & 0 & 0.38(2)& 0.06(1)
\\
 -0.059(5)& 0 & 0 & 0.53(1)&-0.019(8)
\\
 -0.06(1)&-0.02(1)& 0.79(2)& 0 & 0
\\
 -0.053(8)& 0.048(6)&-0.300(9)& 0 & 0
\\
\end{array}
\right)
\]
\[
{\cal Z}^+ =
\left(
\begin{array}{rrrrr}
  0.87(2)& 0 & 0 & 0 & 0
\\
  0 & 0.96(3)& 0.45(1)& 0 & 0
\\
  0 & 0.016(3)& 0.35(1)& 0 & 0
\\
  0 & 0 & 0 & 0.44(1)&-0.009(3)
\\
  0 & 0 & 0 &-0.44(2)& 1.20(3)
\\
\end{array}
\right)
\]
\[
Z^+_\chi =
\left(
\begin{array}{rrrrr}
  0.88(2)& 0 & 0 & 0 & 0
\\
  0 & 0.97(3)& 0.38(1)& 0 & 0
\\
  0 & 0.010(3)& 0.41(1)& 0 & 0
\\
  0 & 0 & 0 & 0.52(2)&-0.009(4)
\\
  0 & 0 & 0 &-0.41(2)& 1.20(4)
\\
\end{array}
\right)
\]
\[
\Delta^+ =
\left(
\begin{array}{rrrrr}
  0 &-0.27(1)&-0.022(7)& 0.14(1)& 0.034(8)
\\
 -0.203(9)& 0 & 0 &-0.49(2)& 0.02(1)
\\
  0.041(2)& 0 & 0 & 0.70(2)& 0.001(7)
\\
  0.029(2)&-0.010(4)& 0.46(1)& 0 & 0
\\
  0.037(5)& 0.012(5)& 0.273(9)& 0 & 0
\\
\end{array}
\right)
\]

\subsection{Wilson action, $\beta=6.0$, $\mu^2 a^2 =0.964$}

\[
{\cal Z}^- =
\left(
\begin{array}{rrrrr}
  0.651(7)& 0 & 0 & 0 & 0
\\
  0 & 0.611(9)& -0.262(5)& 0 & 0
\\
  0 & -0.018(1)& 0.316(7)& 0 & 0
\\
  0 & 0 & 0 & 0.271(8)& 0.007(2)
\\
  0 & 0 & 0 & 0.178(4)& 0.721(9)
\\
\end{array}
\right)
\]
\[
Z^-_\chi =
\left(
\begin{array}{rrrrr}
  0.655(8)& 0 & 0 & 0 & 0
\\
  0 & 0.620(9)& -0.253(4)& 0 & 0
\\
  0 & -0.021(2)& 0.35(1)& 0 & 0
\\
  0 & 0 & 0 & 0.294(9)& 0.008(2)
\\
  0 & 0 & 0 & 0.166(5)& 0.72(1)
\\
\end{array}
\right)
\]
\[
\Delta^- =
\left(
\begin{array}{rrrrr}
  0 &-0.092(3)& 0.014(6)&-0.058(7)&-0.040(7)
\\
 -0.152(8)& 0 & 0 & 0.259(9)& 0.022(6)
\\
 -0.034(4)& 0 & 0 & 0.36(1)& 0.011(4)
\\
 -0.021(4)&-0.023(3) & 0.50(1)& 0 & 0
\\
 -0.024(3)& 0.017(3)&-0.194(6)& 0 & 0
\\
\end{array}
\right)
\]
\[
{\cal Z}^+ =
\left(
\begin{array}{rrrrr}
  0.532(7)& 0 & 0 & 0 & 0
\\
  0 & 0.611(8)& 0.262(5)& 0 & 0
\\
  0 & 0.018(1)& 0.316(8)& 0 & 0
\\
  0 & 0 & 0 & 0.363(7)&-0.015(2)
\\
  0 & 0 & 0 &-0.239(5)& 0.678(9)
\\
\end{array}
\right)
\]
\[
Z^+_\chi =
\left(
\begin{array}{rrrrr}
  0.538(8)& 0 & 0 & 0 & 0
\\
  0 & 0.610(9)& 0.241(6)& 0 & 0
\\
  0 & 0.017(1)& 0.330(8)& 0 & 0
\\
  0 & 0 & 0 & 0.388(8)&-0.016(2)
\\
  0 & 0 & 0 &-0.227(5)& 0.678(9)
\\
\end{array}
\right)
\]
\[
\Delta^+ =
\left(
\begin{array}{rrrrr}
  0 &-0.176(8)&-0.031(5)& 0.054(6)& 0.007(4)
\\
 -0.122(6)& 0 & 0 &-0.30(1)& 0.011(4)
\\
  0.025(1)& 0 & 0 & 0.44(1)&-0.013(2)
\\
  0.011(1)& 0.009(2)& 0.308(9)& 0 & 0
\\
  0.013(2)& 0.010(2)& 0.186(5)& 0 & 0
\\
\end{array}
\right)
\]

\subsection{Clover action, $\beta=6.2$, $\mu^2 a^2 =0.617$}

\[
{\cal Z}^- =
\left(
\begin{array}{rrrrr}
  0.97(2)& 0 & 0 & 0 & 0
\\
  0 & 1.04(2)& -0.51(1)& 0 & 0
\\
  0 & -0.016(2)& 0.337(9)& 0 & 0
\\
  0 & 0 & 0 & 0.28(1)&-0.029(8)
\\
  0 & 0 & 0 & 0.27(1)& 1.21(3)
\\
\end{array}
\right)
\]
\[
Z^-_\chi =
\left(
\begin{array}{rrrrr}
  0.98(3)& 0 & 0 & 0 & 0
\\
  0 & 1.07(2)& -0.49(2) & 0 & 0
\\
  0 & -0.025(3)& 0.48(2) & 0 & 0
\\
  0 & 0 & 0 & 0.32(1)&-0.017(6)
\\
  0 & 0 & 0 & 0.21(1)& 1.24(3)
\\
\end{array}
\right)
\]
\[
\Delta^- =
\left(
\begin{array}{rrrrr}
  0 &-0.14(1)&-0.03(2)&-0.16(1)&-0.11(1)
\\
 -0.20(1)& 0& 0 & 0.37(2)& 0.060(9)
\\
 -0.054(9)& 0 & 0 & 0.52(3)&-0.008(9)
\\
 -0.060(8)&-0.02(1)& 0.77(3)& 0 & 0
\\
 -0.072(4)& 0.042(5)&-0.294(9)& 0 & 0
\\
\end{array}
\right)
\]
\[
{\cal Z}^+ =
\left(
\begin{array}{rrrrr}
  0.94(2)& 0 & 0 & 0 & 0
\\
  0 & 1.04(2)& 0.51(1)& 0 & 0
\\
  0 & 0.017(2)& 0.338(9)& 0 & 0
\\
  0 & 0 & 0 & 0.44(1)&-0.012(3)
\\
  0 & 0 & 0 &-0.52(1)& 1.36(2)
\\
\end{array}
\right)
\]
\[
Z^+_\chi =
\left(
\begin{array}{rrrrr}
  0.96(2)& 0 & 0 & 0 & 0
\\
  0 & 1.03(2)& 0.41(1)& 0 & 0
\\
  0 & 0.007(3)& 0.41(2)& 0 & 0
\\
  0 & 0 & 0 & 0.54(2)&-0.007(4)
\\
  0 & 0 & 0 &-0.47(1)& 1.35(2)
\\
\end{array}
\right)
\]
\[
\Delta^+ =
\left(
\begin{array}{rrrrr}
  0 &-0.25(1)&-0.01(1)& 0.12(1)& 0.040(9)
\\
 -0.200(9)& 0 & 0 &-0.50(2)&-0.004(6)
\\
  0.045(2)& 0 & 0 & 0.70(3)&-0.005(6)
\\
  0.033(2)&-0.011(5)& 0.45(2)& 0 & 0
\\
  0.044(5)& 0.003(5)& 0.27(1)& 0 & 0
\\
\end{array}
\right)
\]

\subsection{Wilson action, $\beta=6.2$, $\mu^2 a^2 =0.617$}

\[
{\cal Z}^- =
\left(
\begin{array}{rrrrr}
  0.72(2)& 0 & 0 & 0 & 0 
\\
  0 & 0.68(1)& -0.31(1)& 0 & 0 
\\
  0 & -0.017(1)& 0.313(7)& 0 & 0 
\\
  0 & 0 & 0 & 0.286(8)& 0.001(3)
\\
  0 & 0 & 0 & 0.21(1)& 0.82(2)
\\
\end{array}
\right)
\]
\[
Z^-_\chi=
\left(
\begin{array}{rrrrr}
  0.72(2)& 0 & 0 & 0 & 0
\\
  0 & 0.69(1)& -0.29(1)& 0 & 0
\\
  0 & -0.021(2)& 0.376(8)& 0 & 0
\\
  0 & 0 & 0 & 0.291(9)& 0.002(3)
\\
  0 & 0 & 0 & 0.17(1)& 0.82(2)
\\
\end{array}
\right)
\]
\[
\Delta^- =
\left(
\begin{array}{rrrrr}
  0 &-0.088(6) & 0.009(5)&-0.063(7)&-0.051(5)
\\
 -0.13(1)& 0 & 0 & 0.27(2)& 0.028(6)
\\
 -0.027(6)& 0 & 0 & 0.40(2)& 0.004(9)
\\
 -0.023(4)&-0.019(4)& 0.55(4)& 0 & 0
\\
 -0.033(5)& 0.022(3)&-0.21(1)& 0 & 0
\\
\end{array}
\right)
\]
\[
{\cal Z}^+ =
\left(
\begin{array}{rrrrr}
  0.60(1)& 0 & 0 & 0 & 0
\\
  0 & 0.68(1)& 0.31(1)& 0 & 0
\\
  0 & 0.017(2)& 0.312(7)& 0 & 0
\\
  0 & 0 & 0 & 0.366(7)&-0.013(2)
\\
  0 & 0 & 0 &-0.287(9)& 0.78(1)
\\
\end{array}
\right)
\]
\[
Z^+_\chi =
\left(
\begin{array}{rrrrr}
  0.60(1)& 0 & 0 & 0 & 0
\\
  0 & 0.68(1)& 0.27(1)& 0 & 0
\\
  0 & 0.015(3)& 0.343(6)& 0 & 0
\\
  0 & 0 & 0 & 0.411(7)&-0.014(2)
\\
  0 & 0 & 0 &-0.27(1)& 0.78(1)
\\
\end{array}
\right)
\]
\[
\Delta^+ =
\left(
\begin{array}{rrrrr}
  0 &-0.155(9)&-0.012(5)& 0.050(6)& 0.024(5)
\\
 -0.118(7)& 0 & 0 &-0.33(2)& 0.006(6)
\\
  0.026(3)& 0 & 0 & 0.49(3)&-0.011(5)
\\
  0.015(2)& 0.004(5)& 0.33(2)& 0 & 0
\\
  0.019(3)& 0.004(4)& 0.190(9)& 0 & 0
\\
\end{array}
\right)
\]

\subsection{Clover action, $\beta=6.4$, ${\mu^2 a^2 =0.313}$}

\[
{\cal Z}^- =
\left(
\begin{array}{rrrrr}
  0.86(2)& 0 & 0 & 0 & 0
\\
  0 & 1.00(2)& -0.50(2)& 0 & 0
\\
  0 & -0.014(4)& 0.308(9)& 0 & 0
\\
  0 & 0 & 0 & 0.229(9)&-0.035(6)
\\
  0 & 0 & 0 & 0.23(1)& 1.16(4)
\\
\end{array}
\right)
\]
\[
Z^-_\chi =
\left(
\begin{array}{rrrrr}
  0.87(2)& 0 & 0 & 0 & 0
\\
  0 & 1.04(2)& -0.51(2)& 0 & 0
\\
  0 & -0.024(5)& 0.39(1)& 0 & 0
\\
  0 & 0 & 0 & 0.27(1) & -0.042(7)
\\
  0 & 0 & 0 & 0.19(1)& 1.17(4)
\\
\end{array}
\right)
\]
\[
\Delta^- =
\left(
\begin{array}{rrrrr}
  0 &-0.116(9)&-0.02(2)&-0.13(2)&-0.11(2)
\\
 -0.17(1)& 0 & 0 & 0.31(1)& 0.02(3)
\\
 -0.044(7)& 0 & 0 & 0.45(2)&-0.04(2)
\\
 -0.05(1)&-0.04(1)& 0.71(4)& 0 & 0
\\
 -0.08(1)& 0.029(8)&-0.24(1)& 0 & 0
\\
\end{array}
\right)
\]
\[
{\cal Z}^+ =
\left(
\begin{array}{rrrrr}
  0.92(1)& 0 & 0 & 0 & 0
\\
  0 & 1.00(2)& 0.50(2)& 0 & 0
\\
  0 & 0.015(2)& 0.31(1)& 0 & 0
\\
  0 & 0 & 0 & 0.406(9)&-0.005(4)
\\
  0 & 0 & 0 &-0.54(2)& 1.37(4)
\\
\end{array}
\right)
\]
\[
Z^+_\chi =
\left(
\begin{array}{rrrrr}
  0.93(1)& 0 & 0 & 0 & 0
\\
  0 & 1.00(2)& 0.42(2)& 0 & 0
\\
  0 & 0.005(4)& 0.34(1)& 0 & 0
\\
  0 & 0 & 0 & 0.47(1)&-0.005(5)
\\
  0 & 0 & 0 &-0.51(2)& 1.37(4)
\\
\end{array}
\right)
\]
\[
\Delta^+=
\left(
\begin{array}{rrrrr}
  0 &-0.21(2)&-0.01(1)& 0.12(2)& 0.04(1)
\\
 -0.18(1)& 0& 0&-0.43(3)& 0.002(19)
\\
  0.040(8)& 0& 0& 0.63(4)& 0.002(11)
\\
  0.027(7)&-0.012(4)& 0.39(2)& 0 & 0
\\
  0.038(9)& 0.008(8)& 0.23(1)& 0 & 0
\\
\end{array}
\right)
\]

\subsection{Wilson action, $\beta=6.4$, $\mu^2 a^2 =0.313$}

\[
{\cal Z}^- =
\left(
\begin{array}{rrrrr}
  0.68(1)& 0 & 0 & 0 & 0
\\
  0 & 0.687(9)& -0.322(9)& 0 & 0
\\
  0 & -0.014(2)& 0.275(5)& 0 & 0
\\
  0 & 0 & 0 & 0.226(7)&-0.009(3)
\\
  0 & 0 & 0 & 0.192(9)& 0.82(3)
\\
\end{array}
\right)
\]
\[
Z^-_\chi=
\left(
\begin{array}{rrrrr}
  0.68(1)& 0 & 0 & 0 & 0
\\
  0 & 0.697(8)& -0.32(1)& 0 & 0
\\
  0 & -0.018(3)& 0.317(9)& 0 & 0
\\
  0 & 0 & 0 & 0.242(7)&-0.010(4)
\\
  0 & 0 & 0 & 0.176(9)& 0.81(3)
\\
\end{array}
\right)
\]
\[
\Delta^- =
\left(
\begin{array}{rrrrr}
  0 &-0.084(4)& 0.012(9)&-0.06(1)&-0.04(1)
\\
 -0.123(6)& 0 & 0 & 0.251(8)& 0.02(1)
\\
 -0.034(3)& 0 & 0 & 0.36(1)&-0.007(9)
\\
 -0.033(5)&-0.025(7)& 0.54(3)& 0 & 0
\\
 -0.025(8)& 0.014(5)&-0.182(8)& 0 & 0
\\
\end{array}
\right)
\]
\[
{\cal Z}^+ =
\left(
\begin{array}{rrrrr}
  0.595(7)& 0 & 0 & 0 & 0
\\
  0 & 0.686(9)& 0.321(8)& 0 & 0
\\
  0 & 0.014(2)& 0.275(6)& 0 & 0
\\
  0 & 0 & 0 & 0.335(5)&-0.011(2)
\\
  0 & 0 & 0 &-0.31(1)& 0.81(2)
\\
\end{array}
\right)
\]
\[
Z^+_\chi =
\left(
\begin{array}{rrrrr}
  0.600(7)& 0 & 0 & 0 & 0
\\
  0 & 0.683(9)& 0.299(9)& 0 & 0
\\
  0 & 0.011(2)& 0.290(7)& 0 & 0
\\
  0 & 0 & 0 & 0.363(7)&-0.012(2)
\\
  0 & 0 & 0 &-0.30(1)& 0.81(2)
\\
\end{array}
\right)
\]
\[
\Delta^+=
\left(
\begin{array}{rrrrr}
  0 &-0.142(9)&-0.012(5)& 0.061(7)& 0.015(6)
\\
 -0.111(5)& 0 & 0 &-0.30(2)& 0.003(10)
\\
  0.029(4)& 0 & 0 & 0.46(2)&-0.007(6)
\\
  0.015(2)& 0.002(3)& 0.30(1)& 0 & 0
\\
  0.019(4)& 0.005(5)& 0.172(8)& 0 & 0
\\
\end{array}
\right)
\]

\newpage
\begin{figure}[t]
\ewxy{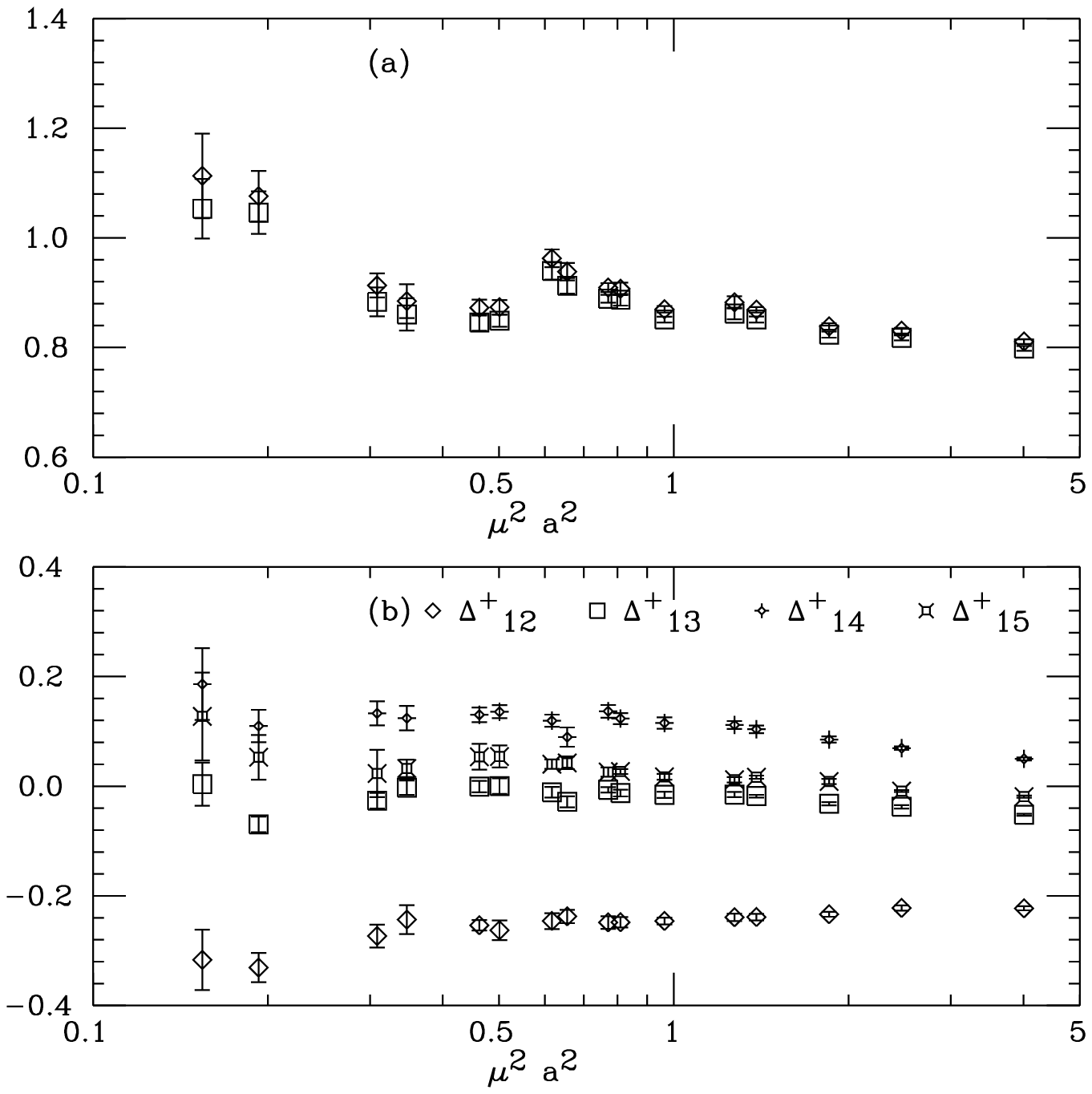}{100mm}   
\vspace{3.0truecm}
%\vspace{0.1cm}
%\centerline{\epsfig{figure=z4fpWILL.ps,height=16cm,angle=0}}
\caption[]{Renormalization constants (at $\beta=6.2$; Clover action)
in the chiral limit as a function of the renormalization scale:
(a) $Z^+_{11}$ ($\diamond$) and $\ct{Z}^+_{11}$ ($\Box$); 
(b) $\Delta^+_{1i},\ i=2,\ldots,5$.}
\protect\label{fig:z4fpWILL}
\end{figure}
\newpage
\begin{figure}[t]   
\vspace{0.1cm}
\centerline{\epsfig{figure=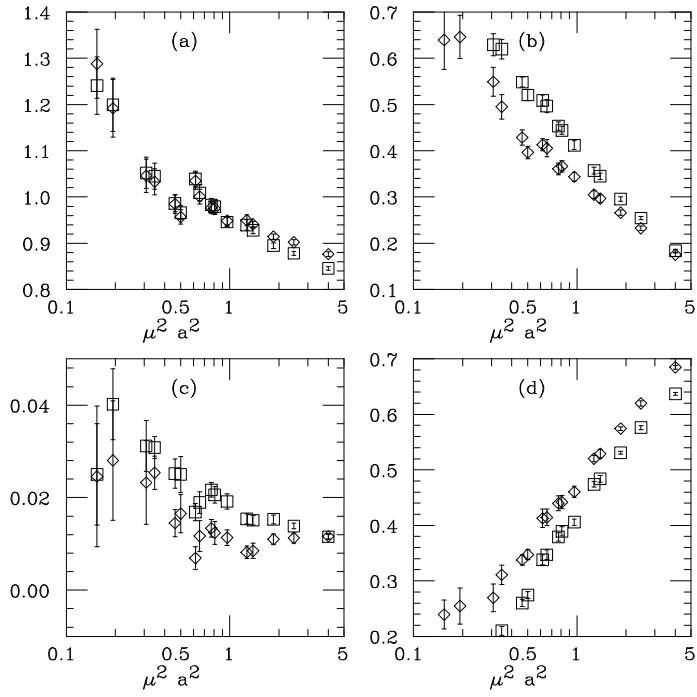,height=16cm,angle=0}}
%\ewxy{z4fpWILR.ps}{100mm}
\caption{Renormalization constants (at $\beta=6.2$; Clover action)
in the chiral limit as a function of the renormalization scale:
(a) $Z^+_{22}$ ($\diamond$) and $\ct{Z}^+_{22}$ ($\Box$); 
(b) $Z^+_{23}$ ($\diamond$) and $\ct{Z}^+_{23}$ ($\Box$); 
(c) $Z^+_{32}$ ($\diamond$) and $\ct{Z}^+_{32}$ ($\Box$); 
(d) $Z^+_{33}$ ($\diamond$) and $\ct{Z}^+_{33}$ ($\Box$).}
\label{fig:z4fpWILR}
\end{figure}
\newpage
\begin{figure}[t]
\ewxy{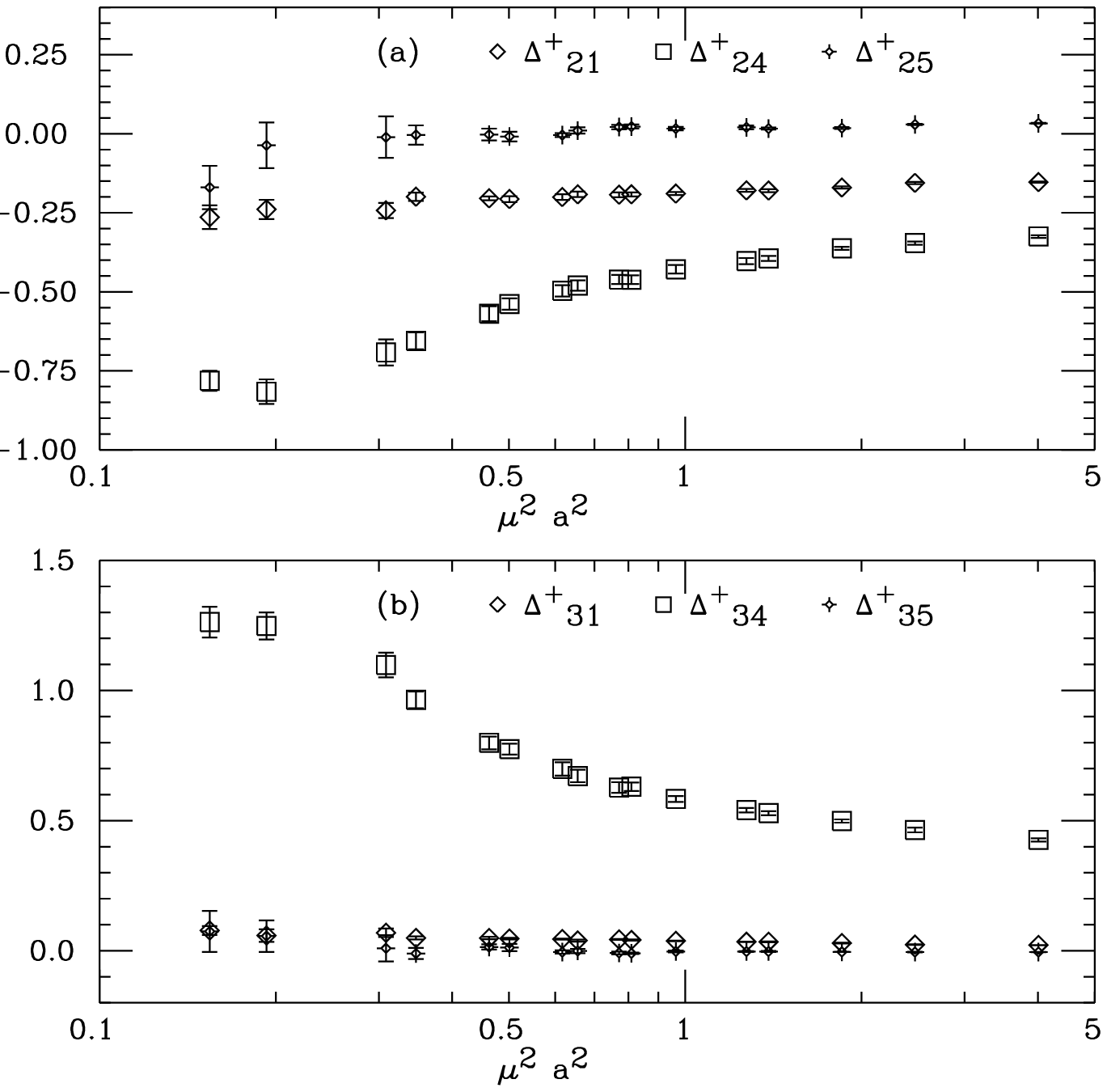}{100mm}
\vspace{3.0truecm}
%\vspace{0.1cm}
%\centerline{\epsfig{figure=z4fpWILRMix.ps,height=16cm,angle=0}}
\caption[]{Renormalization constants (at $\beta=6.2$; Clover action)
in the chiral limit as a function of the renormalization scale:
(a) $\Delta^+_{2i},\ i=1,4,5$; 
(b) $\Delta^+_{3i},\ i=1,4,5$.}
\protect\label{fig:z4fpWILRMix}
\end{figure}
\newpage
\begin{figure}[t]   
\vspace{0.1cm}
%\ewxy{z4fpWITT.ps}{100mm}
\centerline{\epsfig{figure=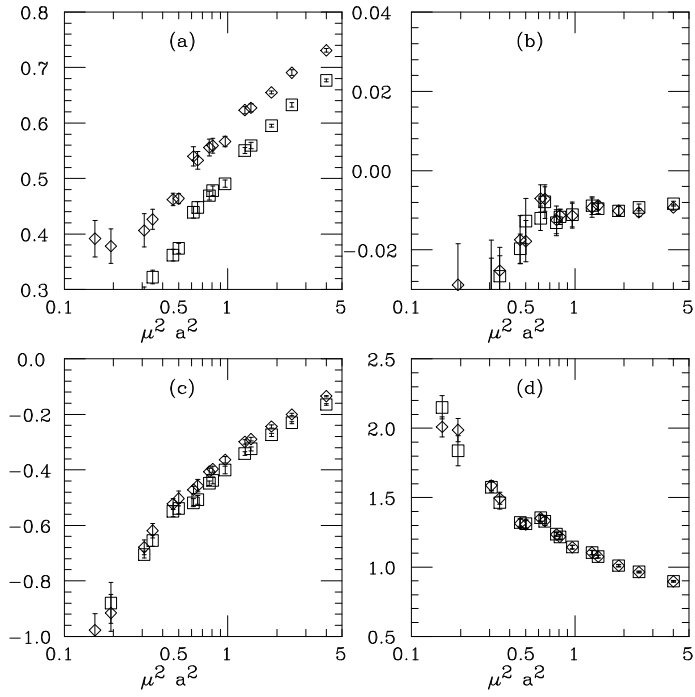,height=16cm,angle=0}}
\caption{Renormalization constants (at $\beta=6.2$; Clover action)
in the chiral limit as a function of the renormalization scale:
(a) $Z^+_{44}$ ($\diamond$) and $\ct{Z}^+_{44}$ ($\Box$);
(b) $Z^+_{45}$ ($\diamond$) and $\ct{Z}^+_{45}$ ($\Box$);
(c) $Z^+_{54}$ ($\diamond$) and $\ct{Z}^+_{54}$ ($\Box$);
(d) $Z^+_{55}$ ($\diamond$) and $\ct{Z}^+_{55}$ ($\Box$).}
\label{fig:z4fpWITT}
\end{figure}
\newpage
\begin{figure}[t]
\ewxy{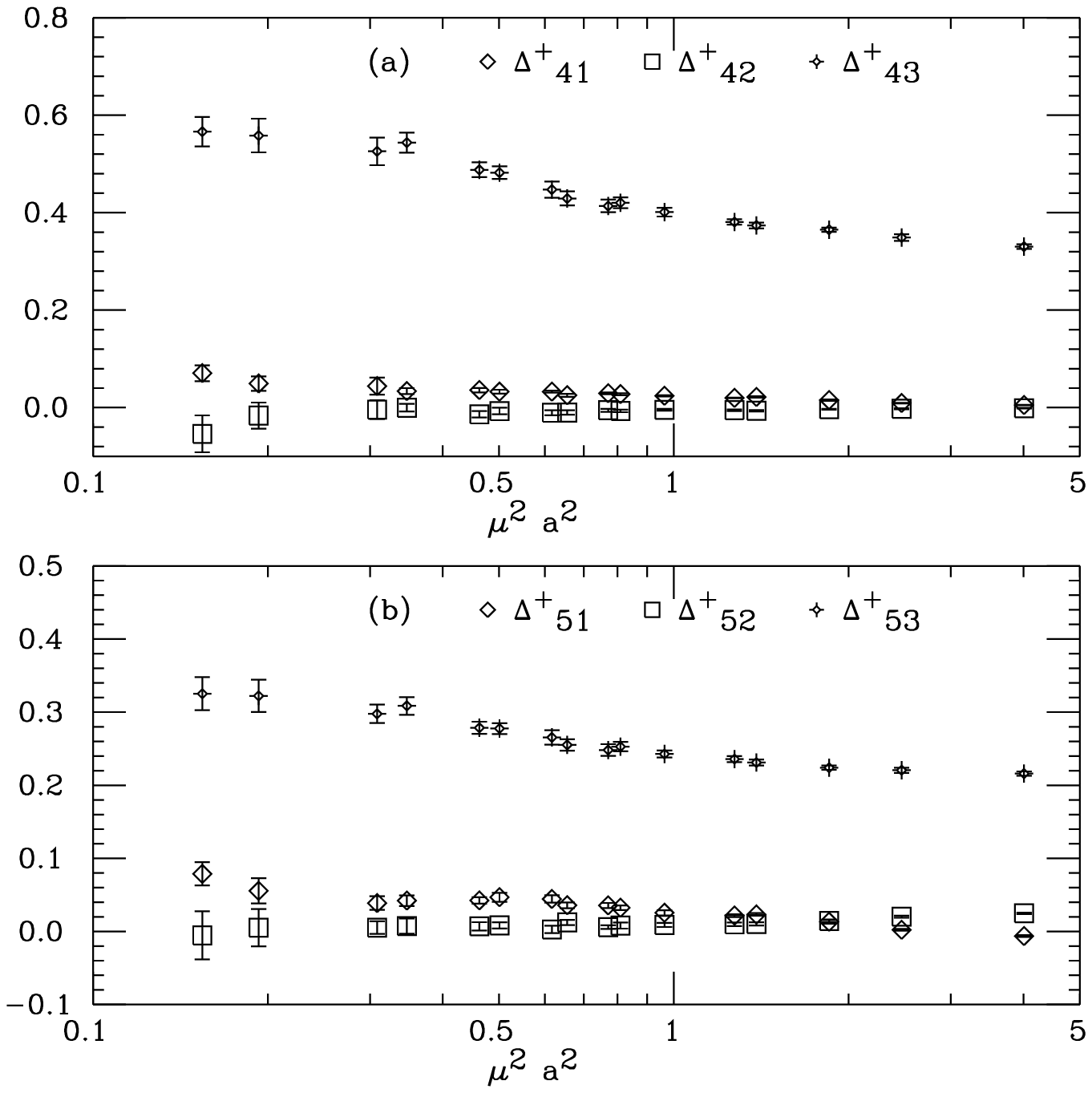}{100mm}   
\vspace{3.0truecm}
%\vspace{0.1cm}
%\centerline{\epsfig{figure=z4fpWITTMix.ps,height=16cm,angle=0}}
\caption[]{Renormalization constants (at $\beta=6.2$; Clover action)
in the chiral limit as a function of the renormalization scale:
(a) $\Delta^+_{4i},\ i=1,2,3$; 
(b) $\Delta^+_{5i},\ i=1,2,3$.}
\protect\label{fig:z4fpWITTMix}
\end{figure}

\end{document}